\documentclass[fleqn,usenatbib]{mnras}

\usepackage{newtxtext,newtxmath}

\usepackage[T1]{fontenc}
\usepackage{multirow}

\usepackage{color} 
\usepackage{soulutf8}

\DeclareRobustCommand{\VAN}[3]{#2}
\let\VANthebibliography\thebibliography
\def\thebibliography{\DeclareRobustCommand{\VAN}[3]{##3}\VANthebibliography}

\usepackage{graphicx}	
\usepackage{amsmath}	
\usepackage{pdflscape}	
\PassOptionsToPackage{hyphens}{url}\usepackage{hyperref}

\title[HYMALAIA]{HYMALAIA: A Hybrid Lagrangian Model for Intrinsic Alignments}

\author[Maion et. al]{
Francisco Maion$^{1,2}$\thanks{E-mail: francisco.maion@dipc.org},
Raul E. Angulo$^{1,3}$,
Thomas Bakx$^{4}$,
Nora Elisa Chisari$^{4}$,
Toshiki Kurita$^{5}$,
\newauthor{Marcos Pellejero-Ibañez$^{1}$}
\\
$^{1}$Donostia International Physics Center, Manuel Lardizabal Ibilbidea, 4, 20018 Donostia, Gipuzkoa, Spain\\
$^{2}$Euskal Herriko Unibertsitatea, Edificio Ignacio Maria Barriola, Plaza Elhuyar, 1, 20018 Donostia-San Sebastián, Spain\\
$^{3}$IKERBASQUE, Basque Foundation for Science, 48013, Bilbao, Spain\\
$^{4}$Institute for Theoretical Physics, Utrecht University, Princetonplein 5, 3584 CC, Utrecht, The Netherlands\\
$^{5}$Kavli Institute for the Physics and Mathematics of the Universe (WPI), The University of Tokyo Institutes for Advanced Study (UTIAS), \\
The University of Tokyo, Chiba 277-8583, Japan
}

\date{Accepted XXX. Received YYY; in original form ZZZ}

\pubyear{2015}

\begin{document}
\label{firstpage}
\pagerange{\pageref{firstpage}--\pageref{lastpage}}
\maketitle

\begin{abstract}
The intrinsic alignment of galaxies is an important ingredient for modelling weak-lensing measurements, and a potentially valuable cosmological and astrophysical signal. In this paper, we present HYMALAIA: a new model to predict the intrinsic alignments of biased tracers. HYMALAIA is based on a perturbative expansion of the statistics of the Lagrangian shapes of objects, which is then advected to Eulerian space using the fully non-linear displacement field obtained from $N$-body simulations. We demonstrate that HYMALAIA is capable of consistently describing monopole and quadrupole of halo shape-shape and matter-shape correlators, and that, without increasing the number of free parameters, it does so more accurately than other perturbatively inspired models such as the non-linear alignment (NLA) model  and the tidal-alignment-tidal-torquing (TATT) model.
\end{abstract}




\section{Introduction}

Light travelling from distant galaxies towards telescopes on Earth is slightly disturbed by the gravitational interaction with the matter distribution in its path \citep{Schneider2006}. This physical process, usually known as weak gravitational lensing, has become a powerful tool for scientific discovery in cosmology \citep[see][for reviews]{Refregier_2003, Hoekstra_2008, MUNSHI_2008, Bartelmann_2010, WEINBERG2013, Kilbinger_2015, Mandelbaum_2018}. Weak lensing is currently one of the most informative cosmological probes available, \citep{HSC_2023_REAL, HSC_2023_SPEC, DES_KIDS_2023, DESY3, Heymans_KiDS_2021, Asgari_KiDS_2021}, and it is bound to increase in importance with the arrival of data from Stage IV experiments such as the Rubin Observatory’s Legacy
Survey of Space and Time (LSST) \citep{thelsstdarkenergysciencecollaboration2021lsst}, \textit{Euclid} \citep{laureijs2011euclid}, or the Nancy Grace Roman Space Telescope (NGRST, formerly WFIRST) \citep{spergel2015widefield}.

The improvement of the quality and quantity of weak-lensing measurements poses the challenge of improving theoretical models and controlling systematical errors, among which can be mentioned photometric redshift errors \citep{salvato_2019, Fischbacher_2023}, baryonic effects \citep{HUTERER_2005, van_Daalen_2011, Hearin_2012, Eifler_2015, Chisari_2018, Huang_2019, Chisari:2019tus, Arico_2023}, and intrinsic alignments (IA) \citep{Joachimi_2015, Kiessling_2015, Kirk_2015}. The latter refers to the fact that the shape of galaxies can be influenced by the matter distribution in their vicinity (e.g. elliptical galaxies tend to align with the principal axis of the local matter tidal tensor), a process by which spatial correlations between these shapes are imprinted. As a consequence, the measurement of galaxy shapes is no longer an unbiased estimator of the cosmic shear, but of its sum with the intrinsic shear. Additionally, IA will also be an important source of systematic errors in spectroscopic surveys, such as DESI, due to target-selection bias \citep{Hirata_2009, Lamman_2023}. Having a reliable way to separate IA from the cosmological signal of interest will be essential to get robust constraints on cosmological parameters.

Besides its importance as a systematic effect, the intrinsic alignments of galaxies can be viewed as an interesting cosmological observable by itself. One can explore its sensitivity to specific physical processes \citep{Kogai_2018, Harvey_2021}, but also use it as an independent cosmological probe \citep{Schmidt_2015, Chisari_2016, Akitsu_2021b, Tsaprazi_2022, Okumura_2022, Kurita_2023, van_Dompseler_2023}, which is expected to provide complementary information to more conventional galaxy clustering statistics, making constraints significantly stronger \citep{Taruya_2020, Okumura_2023}. A full exploration of this potential, however, requires fast and accurate modelling of the signal, with a minimal number of free parameters, so as not to dilute the cosmological information present in the data.

Current models developed to describe IA can be separated into two classes. On the more complex side are models which assign shapes and orientations to galaxies placed into gravity-only simulations, which we will refer to as \textit{semi-analytic} models \citep{Heavens_2000, Heymans_2006, Joachimi_2013_A, Joachimi_2013_B, Hoffmann_2022, vanalfen2023empirical}; these have the advantage of being physically motivated, but the disadvantage of being computationally costly. Moreover, these models must make assumptions regarding the processes through which alignments are imprinted onto galaxies, that may or may not give an accurate description of the physics at play in the real Universe. In an attempt at being insensitive to details of the galaxy formation processes, models based on perturbation theory (PT) \citep{Catelan_2001, Hirata_2004, Bridle_2007, Blazek_2019, Vlah_2020, Vlah_2021, Bakx_2023} or the halo model \citep{Fortuna_2020, Mahony_2022} have also been developed. These have the advantages of being computationally cheap, and being applicable regardless of the specifics of galaxy formation physics; on the downside, their constraining power can depend strongly on the priors given for their free parameters, creating a need for calibration of these priors from simulations and observations. Furthermore, their range of application may be limited, with PT breaking down at small scales when structure formation becomes non-linear, and the halo-model being inaccurate at the transition between the 2-halo and 1-halo terms.

In the case of galaxy clustering, recent works have proposed a way to overcome this limitation of perturbatively inspired models, while retaining the agnostic position relative to galaxy formation. Through the so-called \textit{hybrid Lagrangian bias} approach, one combines the perturbative bias expansion in Lagrangian space with the fully non-linear displacement field from $N$-body simulations, to obtain a model in Eulerian space which is not subject to the limitations of PT in predicting the non-linear evolution of the dark-matter fluid \citep{Modi_2020}. This method has been shown to describe the power spectrum of galaxies in real and redshift space well beyond the limits of PT methods \citep{Zennaro_2023,Kokron_2021,Pellejero_Iba_ez_2022}, for very different assumptions on the galaxy formation model \citep{Zennaro_2022, Kokron_2022}, and it has been applied to real and simulated data to obtain unbiased cosmological constraints \citep{Hadzhiyska_2021, Pellejero_Iba_ez_2023}; it has also been shown to describe well statistics beyond the 2-point function, such as k-nearest neighbour cumulative distribution functions (kNN-CDFs) \citep{Banerjee_2022}. Such results provide strong motivation for applying the hybrid Lagrangian bias method to the modelling of IA: one expects to obtain an accurate and agnostic model that extends well beyond the limits of typical PT models. 

In this work, we construct such a model. We develop a model based on a Lagrangian bias expansion of shapes, which is then advected to Eulerian space using fully non-linear displacements from $N$-body simulations; we refer to this model as HYMALAIA -- an acronym standing for HYbrid Model Advected from LAgrangian space for Intrinsic Alignments. Besides presenting the model, we test it against measurements of shape power-spectra of halos extracted from $N$-Body simulations, evaluating qualitatively and quantitatively the accuracy with which the model describes them. Tests with measurements of the shape power-spectra of galaxies are deferred to future work.

This article is structured as follows. In section \ref{sec:expansion} we describe the bias expansion formalism; in section \ref{sec:Model} we describe our hybrid model, as well as previously existing ones which we will employ for comparison; in section \ref{sec:simulations} we give a brief account of the simulations used for building our model and for comparing models between each other; section \ref{sec:methods} gives an account of the methods we employ to quantitatively compare the performance of different models; we present our results in section \ref{sec:results} and conclude in section \ref{sec:conclusions}.

\section{Bias Expansion}
\label{sec:expansion}
 
Modern theories of structure formation postulate that the abundance of any large-scale structure (LSS) tracer can be written as a combination of three parts,
\begin{equation}
    \delta_s(\mathbf{x})= (\delta_s)_{\rm{loc}}(\mathbf{x})+(\delta_s)_{\rm{h.d.}}(\mathbf{x})+(\delta_s)_{\rm{stoch}}(\mathbf{x}),
    \label{eq:loc_hd_stoch}
\end{equation}
and the subscript $s$ is to emphasize that so far we are working with scalar quantities. The first term, usually referred to as the \textit{local} part can be described as a linear combination of operators built from the matter field:
\begin{equation}
    \begin{split}
        (\delta_s)_{\rm{loc}}& = \sum_i b_i \mathcal{O}_i\\
    \end{split}
    \label{eq:loc_bias_expansion}
\end{equation}
Here, the constants $b_i$ encapsulate the complex information coming from the formation process of each tracer, and quantify the response of the tracer density to changes in the basis operators (see \citealt{Desjacques_2018} and references therein). This form is valid both in Lagrangian and Eulerian space, with the caveat that the bias parameters will have different meanings. The operators entering this expansion will have to satisfy the equivalence principle, be compatible with homogeneity and isotropy of the LSS, and have the correct symmetry properties (i.e. if we wish to perform this expansion for a trace-free rank 2 tensor, then all the operators must also be trace-free rank 2 tensors). Counting the powers of the linear matter density appearing in each of these operators, one can then order them perturbatively; at a fixed order in perturbation theory, there is a finite number of operators contributing to equation (\ref{eq:loc_bias_expansion}), which can be explicitly written down. Retaining terms up to second order, this expansion reads
\begin{equation}
    (\delta_s)_{\rm{loc}}= b_1\delta + b_2\delta^2 + b_{s^2}s^2 + \mathcal{O}(3) + \dots,
\end{equation}
in which $s^2$ is defined by
\begin{equation}
    s^2 = s_{ij}s_{ij},
\end{equation}
and $s_{ij}$ is the traceless tidal field
\begin{equation}
    s_{ij} = \left( \frac{\partial_i\partial_j}{\nabla^2} - \frac{1}{3}\delta^K_{ij} \right) \delta.
\end{equation}

The second term in equation \eqref{eq:loc_hd_stoch}, usually referred to as the \textit{higher-derivative} part, captures the dependence of the tracer density on non-local effects such as baryonic feedback \citep{Lewandowski_2015}, or simply the collapse of matter from a typical region of radius $R_*$, usually denoted as the non-locality scale. These terms involve spatial derivatives of local gravitational observables, and hence are suppressed at large scales by $(kR_*)^{2n}$, which becomes a new expansion parameter. At order $(1+1)$ in $\delta_{L}$ and $(kR_*)^2$, one can derive the operator contributing to the expansion, given by 
\begin{equation}
    (\delta_s)_{\rm{h.d.}} = b_{\nabla^2}\nabla^2 \delta
    \label{eq:hdterm}
\end{equation}
where $\delta$ is the dark matter overdensity.

Finally, the third term in \eqref{eq:loc_hd_stoch} represents stochastic contributions. These have to be taken into account in order to model effects that small scales have on large, perturbative scales. Here, we will consider only one such term, namely  
\begin{equation}
    (\delta_s)_{\rm{stoch}} = \varepsilon.
\end{equation}
The stochastic field $\varepsilon$ is uncorrelated with the dark matter density field and has vanishing expectation value. Again, it should have the same tensorial symmetry properties as the quantity for which we are performing the expansion, i.e., it would be a symmetric trace-free tensor field $\varepsilon_{ij}$ if the tracer field also has these properties. For the case of a scalar, its two-point correlator in Fourier space takes the form 
\begin{equation}
    \langle \varepsilon(\mathbf{k})\varepsilon(\mathbf{k}') \rangle = (2\pi)^3 \delta(\mathbf{k+k'})(A_1+ A_2(kR_*)^2 + \dots )
    \label{eq:stochastic_exp}
\end{equation}
where $A_i$ are free parameters which cannot be predicted from perturbative techniques.

\subsection{Eulerian Shape Expansion}
\label{sec:eul_shape_exp}

Although this formalism has been employed most commonly to expand the density of biased tracers, it has recently been generalized to allow its application in describing tensorial quantities such as the shape field \citep{Schmitz_2018, Blazek_2019, Vlah_2020, Taruya_2021}. Let $S_{ij}$ denote the $3$-dimensional shape field, and we define the normalized trace-free part of it as
\begin{equation}
    g_{ij}(\mathbf{x}) = \frac{1}{S_0}\left( S_{ij}(\mathbf{x}) - \frac{1}{3}S_0(\mathbf{x})\delta^K_{ij} \right),    
\end{equation}
in which $S_0 = \sum_{i=1}^3 S_{ii}$, the trace of the shape tensor. In this case, they find the expression for the second-order bias expansion of galaxy shapes to be
\begin{equation}
    \begin{split}
        g_{ij}(\mathbf{x})\approx \left( c_s + c_{\delta s}\delta \right)s_{ij}(\mathbf{x}) + c_{s\otimes s} & (s\otimes s)_{ij}(\mathbf{x}) + c_t t_{ij}(\mathbf{x}),\\
        & + c_{\nabla^2} \nabla^2 s_{ij}(\mathbf{x}) + \varepsilon_{ij}(\mathbf{x})
    \end{split}
    \label{eq:bias_expansion}
\end{equation}
in which the operators $(s\otimes s)_{ij}, t_{ij}$ can be defined as
\begin{equation}
    \begin{split}
        (s \otimes s)_{ij}(\mathbf{x}) & = \left( s_{ik}s_{kj}(\mathbf{x}) - \frac{s^2(\mathbf{x})}{3}\delta^K_{ij} \right)\\
        t_{ij}(\mathbf{x}) & = \left( \frac{\partial_i\partial_j}{\nabla^2} - \frac{1}{3}\delta_{ij}^K \right)\left( \theta(\mathbf{x}) - \delta(\mathbf{x}) \right),
    \end{split}
    \label{eq:operators_def}
\end{equation}
where $\theta$ is the dimensionless velocity divergence, $\theta=-(\partial_i v_i)/(faH)$. Notice that we denote density bias parameters by the letter $b$ with a subscript indicating which operator it accompanies; for the shape expansion we denote the bias parameters by the letter $c$, with a subscript indicating what is the operator they accompany, in accordance to the notation of \cite{Schmitz_2018}.

\subsection{Lagrangian Shape Expansion}
\label{sec:lag_shape_exp}

The key assumption behind our work is that the shape field will evolve from Lagrangian to Eulerian space only by having its amplitude rescaled by the local volume contraction or expansion, but without suffering any shear (see equation 2.3 of \cite{Chen_2024}). This can be expressed by the equation
\begin{equation}
    g_{ij}(\mathbf{q}) d^3\mathbf{q} = g_{ij}(\mathbf{x})d^3\mathbf{x},
    \label{eq:shape_conservation}
\end{equation}
meaning that the functional shape of the bias expansion in Lagrangian space is the same as that in Eulerian space,
\begin{equation}
    \begin{split}
        g_{ij}(\mathbf{q})\approx \left( c_s + c_{\delta s}\delta \right)s_{ij}(\mathbf{q}) + & c_{s\otimes s} (s\otimes s)_{ij}(\mathbf{q}) + c_t t_{ij}(\mathbf{q})\\
        & + c_{\nabla^2} \nabla^2 s_{ij}(\mathbf{q}) + \varepsilon_{ij}(\mathbf{q}),
    \end{split}
    \label{eq:lag_bias_expansion}
\end{equation}
with the difference that now all the operators are computed in Lagrangian space, and therefore the bias parameters have a different physical meaning from the Eulerian ones \citep{Schmitz_2018, Taruya_2021}. For the purpose of analyzing actual measurements one needs predictions in Eulerian space, and these can be obtained from $g_{ij}(\mathbf{q})$ through the equation
\begin{equation}
    g_{ij}(\mathbf{x}) = \int d^3\mathbf{q} \delta^D\left( \mathbf{x}-\mathbf{q}-\mathbf{\psi}(\mathbf{q}) \right) g_{ij}(\mathbf{q}),
    \label{eq:advect}
\end{equation}
which is simply a reflection of the conservation law in Eq. (\ref{eq:shape_conservation}); in this expression, $\mathbf{\psi}$ is the displacement field, obtained through any calculation method available (e.g.: perturbation theory or simulations). Our goal is to test how well such a model will be able to reproduce realistic shape correlations of haloes, thus implicitly testing the assumptions made earlier.

\subsection{Density Weighting}
\label{sec:dens_weighting}

Notice, that even though we have written $g_{ij}(\mathbf{q})$ as if it was a well defined field for each point in Lagrangian space, in reality we can only measure the tracer ellipticities at the positions where tracers are available. Hence, one should define a continuous tracer-shape field in Lagrangian space by
\begin{equation}
    \Tilde{g}_{ij}(\mathbf{q}) = \frac{1}{\bar{n}} \sum_{\alpha} g_{ij}(\mathbf{q}_{\alpha}) \delta^D(\mathbf{q}-\mathbf{q}_{\alpha}),
\end{equation}
in which the sum over $\alpha$ is in fact over all objects available, making it clear that $\Tilde{g}$ is in fact the density-weighted reduced shape field \citep{Blazek_2015}. Let us assume a halo population with density field 1+$\delta_h$ and reduced-shape field $g^h_{ij}$, then the density-weighted reduced shape field is given by
\begin{equation}
    \Tilde{g} = (1+\delta_h)g^h_{ij}.
\end{equation}
Let us then expand the halo density to first order, $\delta_h \approx b_1 \delta_L$, and we can write the relevant bias expansion keeping only up to second order terms
\begin{equation}
    \begin{split}
        \Tilde{g} = (1+\delta_h)g^h_{ij} \approx c_s s_{ij} + \left(c_{\delta s} + c_s b_1\right) \delta s_{ij} + c_{s\otimes s}& (s\otimes s)_{ij} + c_t t_{ij} \\
        & + c_{\nabla^2} \nabla^2s_{ij}.
    \end{split}
\end{equation}
Notice that even if we were to retain higher order terms in the density bias expansion, they would not contribute if one chooses to keep only up to second order terms in the final, density-weighted, reduced shape bias expansion. We also discard the term $s_{ij} \nabla^2\delta $ since it contributes at order $(2+1)$ in $\delta_L$ and $(kR_*)^2$ respectively, so that we effectively consider it to be a third-order term.

Interestingly, this shows that even if only the linear term is maintained in the shape expansion, density-weighting generates a term which is of the form $\delta s_{ij}$,
\begin{equation}
    (1+b_1\delta)c_s s_{ij} = c_s s_{ij} + b_1c_s \delta s_{ij},
    \label{eq:gen_s_delta}
\end{equation}
and we can redefine the bias parameter associated to this operator as
\begin{equation}
    \tilde{c}_{s\delta} = c_{s\delta} + b_1c_s
\end{equation}
The functional form of this term is fully expected, and was already included in the second-order expansion for the reduced shape. However, taking equation \eqref{eq:gen_s_delta} as a working hypothesis, one can then test explicitly whether the recovered value of $\tilde{c}_{s\delta}$ is compatible with $b_1 c_s$, which would be an indication that the linear term for the shape expansion is sufficient.

\section{Models}
\label{sec:Model}

In this section we will describe the different models which will then be used to fit the measured shape power spectra. The following subsection will be dedicated to establishing useful definitions and notation, before we actually plunge into the description of the different models.

\subsection{Basic Definitions}

To study the IA signal we will decompose the reduced shape tensor $g_{ij}$ into two measures of ellipticity, equivalent to the Stokes' parameters in the study of light polarization,
\begin{equation}
    \begin{split}
        \epsilon_1(\mathbf{x}) & = \frac{1}{2}\Big(g_{11}(\mathbf{x}) - g_{22}(\mathbf{x}) \Big) \\
        \epsilon_2(\mathbf{x}) & = g_{21}(\mathbf{x}),
    \end{split}
    \label{eq:proj}
\end{equation}
but these definitions still have the disadvantage of being dependent on the particular coordinate system chosen. Fourier transforming these quantities one can define the following combinations \citep{Crittenden_2002},
\begin{equation}
    \begin{split}
        E(\mathbf{k}) & = \epsilon_1 (\mathbf{k})\cos(2\phi_k) + \epsilon_2 (\mathbf{k})\sin(2\phi_k)\\
        B(\mathbf{k}) & = -\epsilon_1(\mathbf{k})\sin(2\phi_k) + \epsilon_2(\mathbf{k})\cos(2\phi_k),\\
    \end{split}
    \label{eq:E-B}
\end{equation}
in which $\phi_k$ is the angle of the $\hat{\mathbf{k}}$ vector in the 2D projected plane. Assuming the line of sight to be given by $\hat{\mathbf{z}}$, this can be defined as
\begin{equation}
    \phi_k = \cos^{-1}\left(\frac{k_x}{\sqrt{k_x^2 + k_y^2}}\right).
    \label{eq:phi_ref}
\end{equation}
This decomposition into $E$ and $B$ modes is coordinate independent, and therefore makes it easier to compare among different reported measurements. 

We will generally be interested in computing the auto spectra of the $E$ and $B$ modes, as well as their cross spectra with the density field
\begin{equation}
    \begin{split}
        \langle \delta(\mathbf{k})E(\mathbf{k}') \rangle & = (2\pi)^3 P_{\delta E} (k,\mu) \delta^D(\mathbf{k}+\mathbf{k}') \\
        \langle E(\mathbf{k}) E(\mathbf{k}') \rangle & = (2\pi)^3 P_{EE} (k,\mu) \delta^D(\mathbf{k}+\mathbf{k}') \\
        \langle B(\mathbf{k}) B(\mathbf{k}') \rangle & = (2\pi)^3 P_{BB} (k,\mu) \delta^D(\mathbf{k}+\mathbf{k}'),
    \end{split}
\end{equation}
in which we do not write these expressions for $P_{\delta B}$ and $P_{EB}$ since these spectra are expected to be compatible with zero \citep{Hirata_2004}, as long as no parity-breaking interactions are involved \citep{Biagetti_2020}. Notice that the above spectra are not isotropic, but depend on $\mu$, the cosine of the angle between $\mathbf{k}$ and the line of sight direction. We will generally choose to work with the multipoles of this 2-dimensional power spectrum, given by
\begin{equation}
    P_{XY}^{(\ell)}(k) = \frac{2\ell+1}{2}\int_{-1}^{1} d\mu \mathcal{L}_\ell(\mu) P_{XY}(k,\mu),
\end{equation}
and $\mathcal{L}_\ell$ is the Legendre polynomial of order $\ell$. Cosmic shear will only generate $E$-modes, while IAs generate $B$-modes which will have a non-zero auto power spectrum, $P_{BB}^{(\ell)}(k)$. Fundamentally, though, it will not generate correlation between $E$ and $B$-modes, which therefore can still be used as a tool to mitigate systematic effects in WL measurements.

\subsection{Previously Existing Models}

In this subsection we will describe previously existing models, which will later be used in comparisons to HYMALAIA. 

\subsubsection{Linear Alignment}

We begin by reviewing the linear alignment (LA) model for IA \citep{Catelan_2001, Hirata_2004}. This model assumes that the intrinsic shear of galaxy shapes is proportional to the line-of-sight projected tidal tensor, 
\begin{equation}
    \gamma^I = \frac{c_s}{2} \left( \partial_x^2 - \partial_y^2, 2\partial_x\partial_y \right) \nabla^{-2}\delta,
    \label{eq:linear_alignment}
\end{equation}
therefore, Fourier transforming this shear field one obtains
\begin{equation}
    \gamma^I(\mathbf{k}) = \frac{c_s}{2}\left(\frac{(k_x^2-k_y^2)}{k^2}, \frac{2k_xk_y}{k^2}\right)\delta, 
\end{equation}
and we can combine equations \eqref{eq:E-B} and \eqref{eq:phi_ref} to compute its $E$ and $B$-mode decomposition
\begin{equation}
    \begin{split}
        E(\mathbf{k}) & = \frac{c_s}{2} (1-\mu^2)\delta(\mathbf{k})+\varepsilon(\mathbf{k})\\
        B(\mathbf{k}) & = \varepsilon(\mathbf{k}),
    \end{split}
\end{equation}
leading to the following predictions for the power spectra of $E$- and $B$- modes of the IA field
\begin{equation}
    \begin{split}
        P_{\delta E}(k,\mu) = & \frac{c_s}{2} (1-\mu^2)P_{\rm{lin}}(k)\\
        P_{EE}(k,\mu) = & \frac{c_s^2}{4} (1-\mu^2)^2 P_{\rm{lin}}(k) + A_{SN}\\
        P_{BB}(k,\mu) = & A_{SN},
        \label{eq:pk_linear_alignment}
    \end{split}
\end{equation}
in which $A_{SN}$ is a constant introduced to model the stochastic noise contribution to the auto spectra (cf. equation \eqref{eq:stochastic_exp}).

\subsubsection{Non-Linear Alignment}

When attempting to model the intrinsic shear power spectrum of galaxies, \cite{Hirata_2004} and later \cite{Bridle_2007} noticed that substituting all occurrences of the $P_{\rm{lin}}$ in these expressions by $P_{NL}$, the non-linear matter power spectrum, generally was a better description of the measurements; the model resulting from this prescription is generally known as \textit{non-linear} alignment model (NLA).

\subsubsection{Tidal-Alignment-Tidal-Torquing}
\label{sec:TATT}

The Tidal-Alignment-Tidal-Torquing (TATT) model \citep{Blazek_2019} goes beyond the simple assumption of linear alignment, employing a second order Eulerian bias expansion such as the one in equation \eqref{eq:bias_expansion}, but without the Laplacian term, and neglecting the contribution of the velocity-shear operator. The expression in equation \eqref{eq:bias_expansion} is then evaluated retaining up to second-order contributions, and the relevant correlators are computed using standard perturbation theory techniques; notice that, as pointed out already in \cite{Blazek_2019}, these calculations are not complete up to 1-loop. It is also important to point out that the leading order contributions to this model, which are simply the linear alignment model predictions (see equation \eqref{eq:pk_linear_alignment}), are evaluated by substituting the occurrences of the linear matter power spectrum by the non-linear predictions given by Halofit \citep{Smith_2003, Takahashi_2012}, in a similar way to what is done in the non-linear alignment model. Finally, an important caveat to mention is that the public implementation of TATT available in \verb|FAST-PT| \citep{Blazek_2019, Fang_2017} outputs the predictions for the IA terms in the Limber approximation, which makes the assumption that only modes perpendicular to the line-of-sight  $(\mu=0)$ contribute to the spectra \citep{Limber_1953}; for $P_{\delta E}^{(\ell)}$ the dependence on $\mu$ is separable, and can therefore be included exactly a posteriori. This is not the case for $P_{EE}^{(\ell)}$ and $P_{BB}^{(\ell)}$, and therefore we include an approximate angle dependence, analogous to the one in equation (\ref{eq:pk_linear_alignment}).

We refer the reader to \citep{Blazek_2019, Schmitz_2018} for more details regarding the TATT model, and suffice with enumerating its free parameters at fixed redshift, $c_s, c_{s\delta}, c_{s\otimes s}, A_{SN}$.

\subsubsection{Effective Field Theory}

As a generalization of LA, one can consider the one-loop effective field theory model for intrinsic alignments (EFT of IA) from \cite{Vlah_2020}. It is formulated in the spirit of the Effective Field Theory of Large-Scale Structure (EFT of LSS) (see, e.g.: \citealt{Baumann_2012, Carrasco_2012, Carroll_2014}), which has seen great success in describing the clustering properties of galaxies on mildly nonlinear scales.

The first step is to once again expand the shape field of the relevant biased tracer as a linear combination of operators computed from the matter field, such as in equation \eqref{eq:loc_bias_expansion}. Unlike what was described in sections \ref{sec:eul_shape_exp} and \ref{sec:lag_shape_exp}, we will now have to perform this expansion considering all operators until 3rd order in the linear matter density; this is required so that the computations of the two-point functions are fully consistent, including all 1-loop diagrams. While these operators are mostly independent at the field level, many contributions to the two-point function become degenerate. As a result, the local part of the bias expansion of $g_{ij}$ depends on 6 parameters in total, which we will label $c_s, c_{2,1}, c_{2,2}, c_{2,3}, c_{3,1}, c_{3,2}$. The subscripts serve to indicate that at linear order, only one operator appears with corresponding bias parameter $c_s$, while at second (third) order three (two) new contributions appear.

However, the deterministic part of the bias expansion is in general insufficient to describe the behaviour of the shape field in the mildly non-linear regime. Concretely, one should allow for (i) higher-derivative effects and (ii) stochastic contributions, in similar spirit to the bias models above. Treating the non-locality scale $R$ on a similar footing as the non-linear scale, two new contributions appear, one for the expansion of the density field,
\begin{equation}
    \delta_{\rm{h.d.}}(\mathbf{x}) = b_{\nabla^2} \nabla^2 \delta^{(1)}(\mathbf{x}),
\end{equation}
and another for the expansion of the shape field
\begin{equation}
    (g_{ij})_{\rm{h.d.}}(\mathbf{x}) = c_{\nabla^2} \nabla^2 s_{ij}^{(1)}(\mathbf{x}).
\end{equation}
where the superscript $(1)$ indicates that we consider only the linear density field. Stochastic contributions must also be taken into account in order to model effects that small scales have on large, perturbative scales. 

Altogether, the six multipole power spectra of interest now depend on $9$ bias parameters in total: six from the local expansion of $g_{ij}$, two higher-derivative parameters $b_{\nabla^2}, c_{\nabla^2}$ and one stochastic amplitude $A_{SN}$ for the monopoles of the E- and B-mode auto-spectra, as in equation \eqref{eq:pk_linear_alignment}. In the subsequent comparison, we will also consider a 7-parameter EFT model in which the bias parameters $c_{3,1}, c_{3,2}$ are omitted. This is expected to increase the amount of information on the linear alignment parameter recovered by the EFT model (see \citealt{Bakx_2023}). For explicit expressions of the relevant power spectra and further background on the EFT of IA we also refer to \cite{Bakx_2023}.

\subsection{HYMALAIA}
\label{sec:hymalaia}

Given that one knows the shape field in Lagrangian space by specifying it with the Lagrangian bias expansion\footnote{Notice that in this equation, we have suppressed the velocity-shear operator, $t_{ij}$. We have chosen to do this because keeping it in the expansion seemed to not bring any improvement whatsoever, and made the fits more unstable. The relevance of each of the terms in the expansion will be explored in depth in future works.},
\begin{equation}
    \begin{split}
        g_{ij}(\mathbf{q})\approx \left( c_s + \tilde{c}_{\delta s}\delta \right)s_{ij}(\mathbf{q}) + & c_{s\otimes s} (s\otimes s)_{ij}(\mathbf{q}) \\
        & + c_{\nabla^2} \nabla^2 s_{ij}(\mathbf{q}) + \varepsilon^L_{ij}(\mathbf{q}),
    \end{split}
    \label{eq:hym_bias_expansion}
\end{equation}
we need information on the displacement field $\mathbf{\psi}(\mathbf{q})$ to evaluate equation (\ref{eq:advect}), thus advecting this field to Eulerian space. A comment is in order at this point, regarding the stochastic field introduced in equation \eqref{eq:lag_bias_expansion}. To be fully consistent with equation \eqref{eq:advect} one would have to define the stochastic field $\epsilon^L_{ij}(\mathbf{q})$ in Lagrangian space and then advect it to Eulerian space. It is instructive to calculate the advected stochastic field using the Zeldovich approximation for the displacement field, which gives
\begin{equation}
    \varepsilon_{ij}(\mathbf{k}) = \varepsilon^L_{ij}(\mathbf{k}) + \int \frac{d^3\mathbf{k}_1}{(2\pi)^3}\delta_L(\mathbf{k}_1)\varepsilon^L_{ij}(\mathbf{k}-\mathbf{k}_1) \frac{\mathbf{k}\cdot(\mathbf{k}-\mathbf{k}_1)}{|\mathbf{k}-\mathbf{k}_1|^2},
\end{equation}
and we choose to keep just the lowest order term in this expression. The advection of equation \eqref{eq:hym_bias_expansion} can be obtained from perturbative approaches \citep{rampf2021, Bernardeau:2001qr}, or evaluated directly from $N$-body simulations, obtaining fully non-linear predictions for this field \citep{Modi_2020}.

In building our hybrid model, we will follow the second approach, employing the simulations of volume $(1440\,$Mpc$/h)^3$ described in Table \ref{tab:cosmologies} to perform the advection of the Lagrangian fields. Having advected the fields to Eulerian space, one can then compute their auto and cross power-spectra, and finally linearly combine them with the bias parameters to obtain the full model for the biased tracers. In the following sections we expand on each of these steps.

\subsubsection{Numerical Implementation}

The process which we sketched above can be detailed into a series of steps that we perform to obtain the non-linearly advected Lagrangian operators:
\begin{enumerate}
    \item We begin by computing the linearly evolved Lagrangian density field
    \begin{equation}
        \delta_L(\mathbf{q},z) = \frac{D(z)}{D(z_0)}\delta_L(\mathbf{q},z_0),
    \end{equation}
    and smoothing it at some typical scale $\Lambda\,h/$Mpc, to remove modes with wavenumbers $k>\Lambda$, that are not expected to contribute to the physical process at hand;
    \item We compute the Lagrangian fields
    \begin{equation}
        s_{ij}(\mathbf{q}), \delta s_{ij}(\mathbf{q}), (s\otimes s)_{ij}(\mathbf{q}), \nabla_q^2 s_{ij}(\mathbf{q})
    \end{equation}
    using the smoothed linear density field;
    \item We subtract the Lagrangian positions of particles from their Eulerian positions at the relevant redshift, thus obtaining the displacement field $\mathbf{\psi}_i = \mathbf{x}_i(z) - \mathbf{q}_i$ of the $N$-body simulation;
    \item We advect the quantities from Lagrangian to Eulerian space using the discretized version of equation \eqref{eq:advect},
    \begin{equation}
        \mathcal{O}_{ij}(\mathbf{x}_p) = \sum_{\mathbf{q}\in S_p} \mathcal{O}_{ij}(\mathbf{q}),
    \end{equation}
    and we have denoted by $S_p$ the region in Lagrangian space such that particles end up in the grid cell $\mathbf{x}_p$.
\end{enumerate}

These calculations are performed using the simulations of volume $(1440\,h^{-1}$Mpc$)^3$ described in Table \ref{tab:cosmologies}. These are the largest simulations in the BACCO simulation project, which we will describe briefly in the following section. The Lagrangian fields are computed in a uniform grid of $1080^3$ cells, from a damped linear density field, such that $\delta_L(k>1h/$Mpc$)=0$. This damping is necessary to remove modes which are too small, and thus not expected to contribute physically to the formation of the alignments of galaxies at the scales of interest. The number of cells was chosen to match the number of outputted particles in the simulation; even though it is run with $4320^3$ particles, only one every 64 are actually stored. The advected fields are interpolated to a grid with the same number of cells using a Cloud-in-Cell (CIC) density assignment scheme. The damping scale $k_d=1\,h/$Mpc was chosen so that we only include modes well below the Nyquist frequency $k_{Ny}\approx 2.35\,h/$Mpc, and are thus not subject to aliasing effects \citep{Orzag_1971}.

\subsubsection{Power Spectra}

\begin{figure*}
    \includegraphics[width=\textwidth]{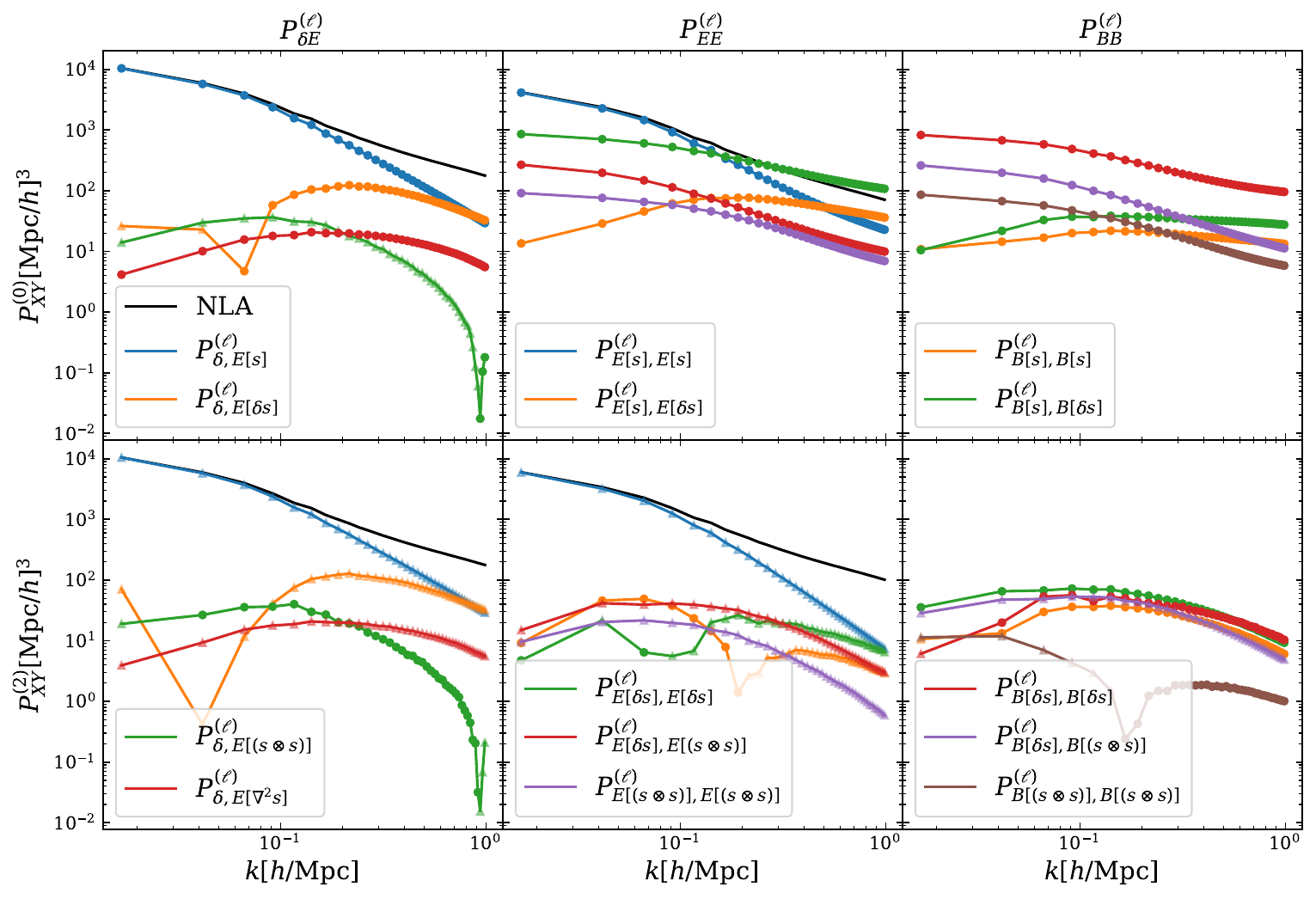}
    \caption{Basis spectra employed in the construction of the HYMALAIA model. Some of these spectra are negative, and hence we show their absolute values, for visualization purposes. The part of these spectra which is positive is shown with circular markers, and the part which is negative is shown with triangular markers in a softer color. The top panel shows the monopole of the auto and cross power-spectra computed between the basis operators entering the bias expansion in equation (\ref{eq:bias_expansion}). Black solid lines show the predictions of NLA for each of the spectra at which they are available; notice that on large-scales they match the lowest order contribution to HYMALAIA. Bottom panel is analogous but displays quadrupole instead of monopole. Not all the spectra that enter the bias expansion are displayed. We have decided to only show the five with largest amplitudes.}
    \label{fig:monopole_compare}
\end{figure*}

Once we have the relevant operators in Eulerian space, $s_{ij}(\mathbf{x})$, $\delta s_{ij}(\mathbf{x})$, $(s\otimes s)_{ij}(\mathbf{x})$, $\nabla_q^2 s_{ij}(\mathbf{x})$, we can build $E$- and $B$-  mode fields for each of these operators, and then compute their auto and cross power spectra. Hence, the model for the shape power spectrum multipoles is given by
\begin{equation}
    \begin{split}
        P^{(\ell)}_{\delta E}(k) & = \sum_{\mathcal{O}\in \left[ s_{ij}, \delta s_{ij}, (s\otimes s)_{ij}, \nabla^2 s_{ij} \right]} c_{\mathcal{O}} P^{(\ell)}_{\delta E_{\mathcal{O}}}(k)\\
        P^{(\ell)}_{XX}(k) & = \sum_{\mathcal{O}, \mathcal{O}' \in \left[ s_{ij}, \cdots, \nabla^2 s_{ij} \right]} c_{\mathcal{O}} c_{\mathcal{O'}} P^{(\ell)}_{X_{\mathcal{O}},X_{\mathcal{O}'}}(k) + A_{SN}\delta^K_{\ell,0},
    \end{split}
\end{equation}
in which $X \in \{E, B\}$, $c_\mathcal{O}$ are the bias parameters associated with the operator $\mathcal{O}$, and $A_{SN}$ is a free constant introduced to model the stochastic noise present in the halo sample. Hence, HYMALAIA at its most complex form has 5 free parameters that need to be adjusted,
\begin{equation}
    c_s, \tilde{c}_{\delta s}, c_{s\otimes s}, c_{\nabla^2} \text{  and  } A_{SN}.
\end{equation}
Figure \ref{fig:monopole_compare} shows some of the basis power spectra entering the equations above; we chose to display only a subset, focusing on those with the largest amplitudes. 

\section{Simulations}
\label{sec:simulations}

The simulations used in this work are summarized in Table \ref{tab:cosmologies}, and are part of the BACCO simulations, described in \citep{Contreras_2020, Angulo_2021, Zennaro_2023}. These are gravity-only simulations run using a modified version of \verb|L-GADGET-3| \citep{Springel_2005, Angulo_2021}. The initial conditions are computed at $z=49$, and the simulation is evolved until $z=0$. At each output redshift we have stored FoF groups and \verb|SUBFIND| subhalos. 

These simulations are run using the \textit{Fixing and Pairing} (F\&P) technique \citep{Angulo_2016}, with the purpose of reducing the variance in the derived $n-$point functions. This procedure consists in generating initial conditions for a simulation without randomly sampling the amplitudes, but doing so only for the phases; the amplitudes instead are all \textit{fixed} to the square root of the power spectrum of interest. One then generates a second set of initial conditions with all phases summed of $\pi$ -- this is equivalent to multiplying all Fourier modes by $-1$; the expected effect is that the deviations of the $n$-point functions from the ensemble average in the two simulations will be anti-correlated, and averaging over their statistics will cancel these deviations exactly. 

The effect of this technique has been characterized for a wide range of observables \citep{Villaescusa_Navarro_2018, Chuang_2019, Klypin_2020, Maion_2022}, but never for the shape power-spectra. Nevertheless, one can obtain rough expectations about the behaviour of the variance of these spectra from results established in these works. At large scales, \cite{Villaescusa_Navarro_2018, Maion_2022} have shown that the variance reduction can be as large as several orders of magnitude, and as small as having roughly no effect at all. On small scales, the scenario is more predictable, since non-linearities in the formation of structures will generally erase any memory of this particular choice of initial conditions, making the power spectra uncorrelated at these scales. If this is the case, then the variance reduction is merely a factor of $2$, the same as one would obtain by running two independent simulations. Another possible scenario observed in \cite{Maion_2022} is that the errors in the paired spectra instead of being anti-correlated are positively correlated. In the limiting case where this correlation coefficient is unity, averaging over these two power spectra will have no variance reduction effect at all, since the power spectra are mathematically the same.

In the remainder of this work we will assume that pairing reduces the variance at all scales by a factor of 2, and will not attempt to model the effect of fixing, but limit ourselves to noticing that values of reduced $\chi^2$ below $1$ at large-scales are most likely due to the variance reduction arising from this procedure. At small scales we can compute the covariance of the shape power spectra using the jacknife technique (see appendix \ref{sec:cov_matrix} for further details), and thus investigate the effect of F\&P. Figure \ref{fig:pairing_jack_cov} shows that for all spectra, with the exception of $P_{EE}^{(0)}$ and $P_{BB}^{(0)}$, the variance is reduced by a factor of 2. For the two exceptional spectra, however, the variance can be reduced by as much as a factor of 5; surprisingly, this effect is seen to increase when going to small-scales, which is opposite to the trends observed in previous works \citep{Angulo_2016, Villaescusa_Navarro_2018, Klypin_2020, Chuang_2019}, and to the theoretical arguments developed in \citep{Angulo_2016, Maion_2022}. This seems to indicate that the shape-noise is highly dependent on the initial phases, and it can be greatly suppressed by pairing.

\begin{table*}
 \caption{Parameters of the simulations used in this work. Notice that for each cosmology, we have two simulations, one smaller, with a volume of $(512\,h^{-1}$Mpc$)^3$, and one larger with volume of $(1440\,h^{-1}$Mpc$)^3$.}
 \begin{tabular}{ccccccccccc}
  \hline 
  & $\sigma_8$ & $\Omega_m$ & $\Omega_b$ & $n_s$ & $h$ & $M_\nu$ & $L[h^{-1}\rm{Mpc}]$ & $m_p [h^{-1} M_{\sun}]$ & $N_p$ & $z$ \\
  \hline
  Nenya & $0.9$ & $0.315$ & $0.05$ & $1.01$ & $0.60$ & $0.0$ & $[512, 1440]$ & $3.2\times 10^9$ & $[1536^3,4320^3]$ & $0$\\
  Narya & $0.9$ & $0.36$ & $0.05$ & $1.01$ & $0.70$ & $0.0$ & $[512, 1440]$ & $3.7\times 10^9$ & $[1536^3,4320^3]$ & $-0.2$\\
  The One & $0.9$ & $0.307$ & $0.05$ & $0.96$ & $0.68$ & $0.0$ & $[512,1440]$ & $3.2\times 10^9$ & $[1536^3,4320^3]$ & $-0.2$\\
  \hline
 \end{tabular}
 \label{tab:cosmologies}
\end{table*}

\subsection{Halo Samples}

We have split the halo population in this simulation into 4 groups selected by mass. Table \ref{tab:mass_samples} displays the information on the selected mass bins and the approximate number density of each type of halo. The smallest halos considered, with mass $M\approx 10^{12}M_{\odot}$, will be composed of approximately $\sim 300$ particles. The mass-definition used for this selection was the one usually referred to as $M_{200,b}$ given by
\begin{equation}
    M_{200,b} = \frac{4\pi}{3} \Delta \rho_b r_{200,b}^3,
\end{equation}
in which $\rho_b$ is the mean matter density of the Universe, and $\Delta=200$; this choice is motivated by the fact that this halo boundary definition is the least affected by changes in cosmology \citep{Ondaro_Mallea_2021}.

\begin{table}
 \caption{Halo samples utilized in this work. These results are quoted for the Nenya cosmology at redshift $z=0.0$. }
 \begin{tabular}{ccc}
  \hline
   & Mass Range $\left[ \log_{10}(M/M_{\sun}) \right]$ & $\bar{n}
   (z=0)\left[h^{-1}\rm{Mpc}\right]^{-3}$ \\
   \hline
   $M_1$ & $\left[ 12, 12.5 \right]$ & $26.9\times 10^{-4}$ \\
   $M_2$ & $\left[ 12.5, 13 \right]$ & $10.1\times 10^{-4}$  \\
   $M_3$ & $\left[ 13, 13.5 \right]$ & $3.7\times 10^{-4}$  \\
   $M_4$ & $\left[ 13.5, 14.5 \right]$ & $1.7\times 10^{-4}$ \\ 
  \hline
 \end{tabular}
 \label{tab:mass_samples}
\end{table}

\subsection{Shape Definition}

For each halo in the simulation, we define its shape through the expression\footnote{This is commonly called the \textit{inertia tensor} in the literature, and denoted by the letter $I$. We decide to keep the notation consistent with other fields of physics, and reserve the name \textit{moment of inertia} for the quantity $I=\sum_{n=1}^N (r_n^2 - x^{(n)}_ix^{(n)}_j)$ , in which $n$ runs through the $N$ particles contained in the relevant object, while the definition in equation (\ref{eq:shape_tensor}) is denoted by the letter $S$ and called \textit{shape tensor}.}
\begin{equation}
    S_{ij} = \sum_{n=1}^N \left(x_i^{(n)} - \overline{x}_i\right)\left(x_j^{(n)} - \overline{x}_j\right),
    \label{eq:shape_tensor}
\end{equation}
for which $\overline{x}_i$ is the mean of $x_i$ for the particles belonging to this specific halo, and $n$ is an index that goes over all halo particles. Many definitions of the halo shape are possible \citep{Bett_2012}, some of them designed to better capture the expected physical effects relevant to galaxy alignments \citep{Tenneti_2014, Kurita_2020}. Nevertheless, in this work we have used the one introduced above because we expect the bias expansion to be general enough to describe measurements with any of the shape definitions (see appendix B of \citealt{Bakx_2023} for cases where this may not hold).

\section{Methods}
\label{sec:methods}

In this section we will describe the methods employed in the analyses to be presented in the following section.

\subsection{Shape Power Spectrum Estimator}

In this section we discuss briefly the methods used to measure the $E$- and $B$- mode shape power spectra of IA \citep{Kurita_2020}. Let $S$ be the shape tensor of a halo, then we can define its ellipticity as \footnote{The ellipticity is defined in the plane perpendicular to the $z$-axis, to be compatible with realistic scenarios where one does not have access to the galaxy shape in the line-of-sight direction.}
\begin{equation}
    \begin{split}
        \epsilon^i_1 & = \frac{1}{2}\frac{S_{xx} - S{yy}}{S_{xx} + S_{yy}}\\
        \epsilon^i_2 & = \frac{S_{xy}}{S_{xx} + S_{yy}},
    \end{split}
    \label{eq:ellipticity_definition}
\end{equation}
and from this, we can formally define a continuous ellipticity field as
\begin{equation}
    \begin{split}
        \epsilon_{1,2}(\mathbf{x}) & =\frac{1}{\bar{n}} \sum_i \epsilon_{1,2}^i \delta^D(\mathbf{x}-\mathbf{x}_i).
    \end{split}
\end{equation}
In practice, we will interpolate the halo ellipticities onto a regular grid of $384^3$ cells, using a cloud-in-cell (CIC) method, giving a Nyquist frequency of $k_{\rm{Ny}}\approx 2.35\,h/$Mpc. This allows us to efficiently evaluate these fields in Fourier space using fast Fourier transforms (FFT); once these fields have been computed in Fourier space we can define the $E$- and $B$-mode fields analogously to equation (\ref{eq:E-B}).

Our estimator for the multipole $\ell$ of the cross power-spectrum between any two fields $X$ and $Y$ is given by
\begin{equation}
    P_{XY}^{(\ell)}(k_i) = \frac{2\ell + 1}{N_i} \sum_{\mathbf{k}\in i\text{th shell} } X(\mathbf{k}) Y(-\mathbf{k}) \mathcal{L}_\ell(\mu(\mathbf{k})),
\end{equation}
in which $\mu=\hat{\mathbf{k}}\cdot \hat{\mathbf{n}}$ is the cosine of the angle between the wavevector $\mathbf{k}$ and the $i$th shell corresponds to the region of a Fourier-space shell with radius in the range $[k_i-\frac{\Delta k}{2}, k_i+\frac{\Delta k}{2}]$.

\subsection{Covariance Matrix}
\label{sec:covariance_matrix}

To model the covariance matrix we will make the assumption that the $E$ and $B$-modes of the shape field are gaussianly distributed, which allows us to write the following analytical expression for the covariance matrix of their auto and cross power-spectra
\begin{equation}
    \begin{split}
        \rm{Cov}&\left(P_{XY}^{(\ell)}(k_i), P_{X'Y'}^{(\ell)'}(k_j)\right) = \frac{(2\ell+1)(2\ell'+1)}{N_i^2}  \delta_{ij} \times \\
        & \sum_{\mathbf{k}\in i\text{th shell}} \mathcal{L}_{\ell}(\mu) \mathcal{L}_{\ell'}(\mu)\left( P_{XX'}(\mathbf{k})P_{YY'}(\mathbf{k}) + P_{X'Y}(\mathbf{k})P_{XY'}(\mathbf{k}) \right),
    \end{split}
    \label{eq:cov_model}
\end{equation}
in which $N_i$ is the number of modes falling inside the $i$th shell. In this expression, we will employ the NLA model to evaluate the power spectra $P_{XY}(\mathbf{k})$; this choice is motivated by Figure \ref{fig:cov_mat_compare}, which displays a direct comparison of LA and NLA to a fully numerical calculation of the covariance matrix performed by \cite{Kurita_2020} from the Dark Quest simulation suite \citep{Nishimichi_2019}. 
Indeed, one can see from Figure \ref{fig:cov_mat_compare} that employing LA greatly underestimates the diagonal terms of the covariance at small scales, while using NLA gives a resonable description for most of the quantities at interest, with the exception of $P_{EE}^{(0)}$ and $P_{BB}^{(0)}$ for which it underestimates the error beyond scales $k \approx 0.5\,h/$Mpc.

To compute this covariance matrix analytically using NLA, we need to determine the value of the free parameters entering the model, namely $c_s$ and $A_{SN}$. $A_{SN}$ can be determined from the small-scale limit of $P_{BB}^{(0)}$, where it should be well approximated by a constant matching the value of the stochastic amplitude. Therefore, we take the bins of this power spectrum with wavemodes between $0.9$ and $1\,h/$Mpc, and compute their average and standard deviation, giving us an estimator of $A_{SN}$ and its error $\sigma_A$. If we assume that at large scales the linear alignment model is valid, then the linear bias $c_s$ can be determined from the expression 
\begin{equation}
    c_s \approx 3\frac{P_{\delta E}^{(0)}}{P_{\rm{lin}}}(k\rightarrow 0),
\end{equation}
as can be seen by inspection of equations \eqref{eq:pk_linear_alignment}, after integrating over $\mu$ to obtain the monopole. Similarly to what is done for $A_{SN}$ we compute the mean and standard deviation of the ratio above for $k<0.05\,h/$Mpc, thus obtaining estimates of $c_s$ and $\sigma_{c_s}$. Even though in the majority of the paper we compute the power spectrum in linearly spaced bins of $k$, to compute this ratio we employ the power spectrum computed in logarithmically spaced bins, so that we can access large scales in a more direct way; the precise values obtained can be seen in Table \ref{tab:fiducial_parameters}.

Once in possession of estimates for $c_s$ and $A_{SN}$, we can evaluate equation \eqref{eq:cov_model} for each of the mass-bins. As stated earlier, this model does not give a good description to the diagonal of the auto-covariances of $P_{EE}^{(0)}$ and $P_{BB}^{(0)}$ at small scales. To account for this, we have numerically estimated the auto-covariance of these spectra at small scales using the jacknife technique. We now parametrize the small-scale behaviour of these covariances for each mass-bin, and add it to the model in equation \eqref{eq:cov_model}. A more detailed account of this procedure can be found in Appendix \ref{sec:cov_matrix}.

Finally, a comment is in order regarding the contribution of model error to the covariance matrix. As described in section \ref{sec:hymalaia}, HYMALAIA is built from numerical simulations and hence contains some level of statistical noise, inherently present due to the finite volume of the simulations. This model noise should in principle be added to the covariance matrix. To avoid having to model this term, we compute the model from simulations with volume $V_L = (1440\,h^{-1}$Mpc$)^3$; the shape power spectra used as test data are instead computed from simulations with volume $V_M = (512\,h^{-1}$Mpc$)^3$. This means that the model noise is highly suppressed with respect to the variance of the data, since the former is computed from a volume approximately 20 times larger than the latter.

\subsection{Performance Metrics}

\subsubsection{Reduced $\chi^2$}

The reduced chi-squared provides us with a quantitative test to assess the quality of the fits. This quantity is defined by
\begin{equation}
    \begin{split}
        \chi^2_{\rm{red}} =\frac{1}{N_{\rm{dof}}} \sum_{\ell,\ell'=0,2} \sum_{\alpha, \beta} \sum_{i,j} \Big( P^{(\ell)}_\alpha(k_i, \Theta) & -  \widehat{P}^{(\ell)}_\alpha(k_i) \Big) \left[C_{\alpha,\beta}^{\ell,\ell'}\right]^{-1}_{ij}\\
        & \Big(P^{(\ell')}_\beta(k_j, \Theta) - \widehat{P}^{(\ell')}_\beta(k_j) \Big),
    \end{split}
\end{equation}
in which $\alpha, \beta$ range between the types of spectra being fitted, i.e. $P_{\delta E}, P_{EE}, P_{BB}$ and $C_{\alpha,\beta}^{\ell,\ell'}$ is the cross-covariance between spectra $P_{\alpha}^\ell$ and $P_{\beta}^{\ell'}$; the number of degrees of freedom is defined by $N_{\rm{dof}} = N_k - N_{\rm{pars}} - 1$, in which $N_k$ is the number of $k$-bins and $N_{\rm{pars}}$ is the number of free parameters being fitted; symbols with a hat represent measured quantities, while those without represent the model being adjusted, and $\Theta$ are its free parameters. 

A comment is in order regarding the interpretation of the values of $\chi^2_{\rm{red}}$ to be presented in this work. We expect these values to be generally below $1$ for the fits to the dataset described above, since the simulations from which the shape power-spectra are computed were run with \textit{Fixed-and-Paired} initial conditions. The effect of this technique is to reduce the variance on statistics derived from these simulations. Since the expression for the covariance matrix defined in equation \eqref{eq:cov_model} assumes Gaussian statistics of the underlying field, it most likely overestimates the variance on the shape power spectra on large scales. As discussed in section \ref{sec:simulations}, the exact effect of this technique on shape power spectra is unknown, but it is reasonable to assume that on small scales the variance will be approximately reduced by half. Hence, to correct for this effect, we use the covariance matrix corresponding to twice the volume of one simulation of $L=512\,h^{-1}$Mpc in computing the values of $\chi^2$. Given the large uncertainties on the actual value of the covariance matrix at small scales, these values of $\chi_{\rm{red}}^2$ should not be interpreted in an absolute manner, but in comparison between different models; a model with a higher value of $\chi_{\rm{red}}^2$ is sure to be providing a worse fit, but it is unclear whether this would indicate the breakdown of that particular approach.

\subsubsection{Figure of Bias}
\label{sec:fob}

The figure of bias provides a consistent way to determine whether the model is fitting the data at the cost of returning spurious bias coefficients, or if there is consistency between the values found while using progressively smaller scales. This quantity can be defined as
\begin{equation}
    \rm{FoB}(k_{\rm{max}}) = \frac{\left|c_s^{\rm{fid}} - c_s(k_{\rm{max}})\right|}{\sqrt{\sigma_{\rm{fid}}^2 + \sigma^2_{c_s}(k_{\rm{max}})}},
\end{equation}
in which $c_{s}^{\rm{fid}}$ and $\sigma_{\rm{fid}}$ are the fiducial linear alignment parameter and the error on the determination of this value, respectively. Notice we have chosen not to include $c_{\delta s}$, $c_{s\otimes s}$ nor $c_{\nabla^2}$ in the calculation of this quantity, and thus will use only $c_s$ in this calculation. This is the only parameter common to all the models considered, and is also the one expected to be degenerate with cosmological parameters and systematic effects. This is supported by Figure \ref{fig:monopole_compare}, in which we see that the contributions proportional to $c_s$ are the ones with the highest amplitude, and thus can easily absorb larger parts of the cosmological signal. The contributions proportional to the higher-order bias parameters, on the other hand, are usually highly suppressed. A secondary reason is that we have not renormalized the operators entering our bias expansion, which is generally necessary to correctly interpret the derived values of the bias parameters \citep{McDonald_2006, Assassi_2014, Werner_2019}. Generally, one would then expect that these \textit{bare} bias parameters be highly scale-dependent, as they absorb higher order contributions that are degenerate with the $n$-point function they multiply. For the case of $c_s$ we will empirically show that it does not suffer from this issue in the range of scales analyzed, and up to the statistical precision with which it is determined.

\subsubsection{Figure of Merit}

The figure of merit (FoM) is a quantitative measure of the amount of information recovered on a chosen set of parameters, and can be defined by the equation
\begin{equation}
    \rm{FoM} = \sqrt{\det\left[ \frac{\Theta_{\alpha\beta}}{\theta^{\rm{fid}}_\alpha\theta^{\rm{fid}}_\beta} \right]^{-1} },
\end{equation}
in which $\Theta$ is the parameter covariance. In this work, however, we will use only $c_s$ in the calculation of this quantity, so that the expression reduces to
\begin{equation}
    \rm{FoM} = c_s/ \sigma_{c_s}
\end{equation}
in which $\sigma_{c_s}$ is computed after marginalizing over the remaining free parameters. Analogously to what was argued in \ref{sec:fob} we do this because $c_s$ is the parameter expected to be most correlated with cosmological parameters.

\subsection{Fitting Procedure}

We employ the \verb|MultiNest| algorithm \citep{Feroz_2008, Feroz_2009, Feroz_2019} through its python interface \verb|pymultinest| \citep{Buchner_2014}
to obtain the best-fitting parameters for each of the models, and explore the multi-dimensional likelihood to obtain confidence regions for each of the parameters of interest.

As in every Bayesian inference method, one must define priors for the free parameters in each explored model. Table \ref{tab:priors} displays the ranges for each free parameter in each of the tested models. These prior ranges are chosen so that they are large enough not to have a noticeable effect on the constraints, that is, are uninformative.

\section{Results}
\label{sec:results}

In this section, we present the tests made to validate the capability of the model to describe shape power spectra and extract reliable values of the IA biases. These shape power spectra are computed from the set of simulations described in section \ref{sec:simulations}. The validation results will be presented using the results obtained with the \textit{Nenya} simulations at redshift $z=0.0$. We have chosen to display these results because this halo sample has a larger value of the intrinsic-alignment bias parameter, and thus represents a more interesting test to the model, with a larger signal-to-noise ratio.

\subsection{Validation}

\begin{figure}
    \includegraphics[width=\columnwidth]{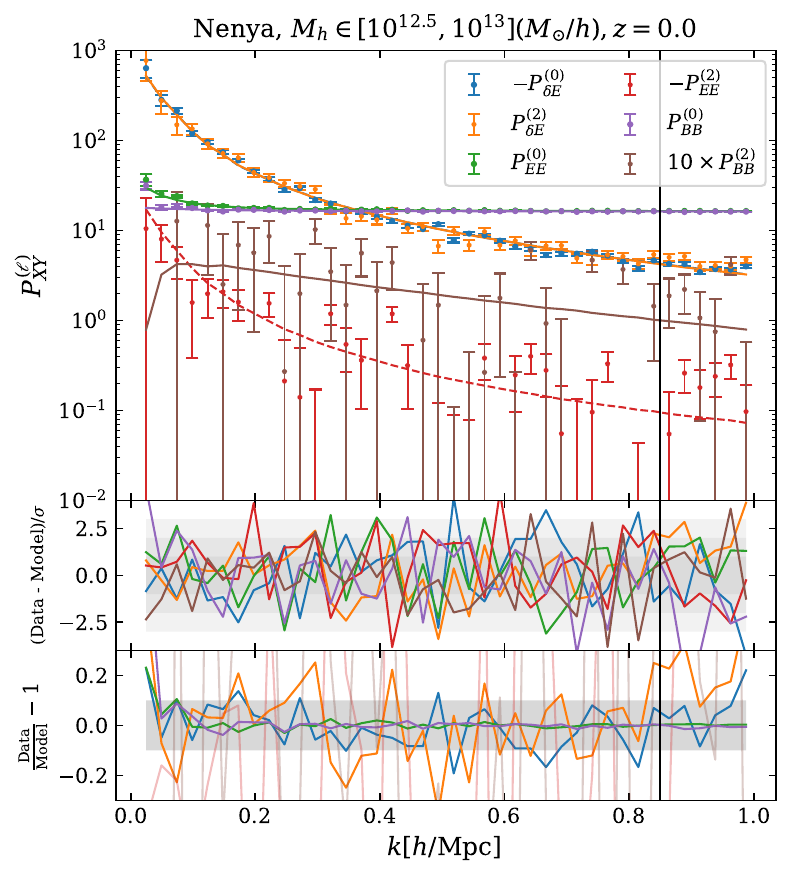}
    \caption{\textit{Upper Panel} Points with error bars indicate the shape power spectrum multipoles measured from the simulations with Nenya cosmology, $L=512\,h^{-1}$Mpc, at redshift $z=0$. Colored solid lines represent the fits using HYMALAIA. Dashed lines indicate spectra that are originally negative, but are plotted here with the reversed sign for visualization purposes. The black vertical line marks the maximum scale used in the fit, $k_{\mathrm{max}}= 0.85\,h/$Mpc. This scale was chosen since it is the smallest scale we ever use in our fits; this is done to avoid entering a regime which could be afected by the damping of modes $k>1\,h/$Mpc made to the linear density fields when building HYMALAIA. \textit{Middle Panel} Colored lines represent difference between model and data, in units of $\sigma$; gray-shaded regions indicate 1, 2 and 3-$\sigma$ regions. \textit{Lower Panel} Shows the fractional difference between model and data. We give special emphasis to these results for $P_{\delta E}^{(0, 2)}$, $P_{EE}^{(0)}$ and $P_{BB}^{(0)}$, since these are expected to be the most relevant terms in a realistic 3x2pt analysis; the gray-shaded band shows $10\%$ deviations. }
    \label{fig:fit_halo-shape_pk}
\end{figure}

The models presented in section (\ref{sec:Model}) were used to fit simultaneously the power-spectrum multipoles $\left[ P_{\delta E}^{(0)}, P_{\delta E}^{(2)}, P_{EE}^{(0)}, P_{EE}^{(2)}, P_{BB}^{(0)}, P_{BB}^{(2)}\right]$ computed from the halo samples described above. An example can be seen in Figure \ref{fig:fit_halo-shape_pk}, which displays a comparison between measurements and HYMALAIA evaluated at the set of best-fitting parameters for an intermediate-mass bin in the Nenya simulation at $z=0.0$. We have chosen to display the result for the mass bin $\log_{10}(M_h/M_{\odot}) \in [12.5, 13]$, labelled $M_2$ in Table \ref{tab:mass_samples}, because we expect halos at this mass to be the ones typically populated by galaxies targeted by Stage IV weak lensing surveys such as LSST and \textit{Euclid} \citep{Korytov_2019}. The deviations of the model from the measurements are mostly below 3-$\sigma$ and usually below 2-$\sigma$ until the maximum scale used in the fit $k_{\rm{max}}=0.85\,h/$Mpc, marked by a vertical black line, demonstrating good agreement.

It is also important to contextualize the accuracy of our model, comparing it to the requirements imposed by the precision of Stage-IV survey measurements. \cite{Paopiamsap_2024} found that keeping mis-modellings below the level of 10$\%$ in the IA contribution to cosmic-shear is sufficient to recover unbiased constraints on cosmology. The lower panel of Figure \ref{fig:fit_halo-shape_pk} shows that our models for $P_{\delta E}^{(0)}$, $P_{EE}^{(0)}$ and $P_{BB}^{(0)}$ are good descriptions of the measured spectra, with deviations at the 10$\%$ level in the case of $P_{\delta E}^{(0)}$, and well beyond that threshold for the other two spectra. These spectra are the ones expected to contribute most relevantly to the cosmic-shear signal and therefore, we can conclude that our simulations can test the intrinsic-alignment contribution with a similar precision to that of Stage-IV surveys.
 
\subsubsection{Mass-Samples}

\begin{figure}
	\includegraphics[width=\columnwidth]{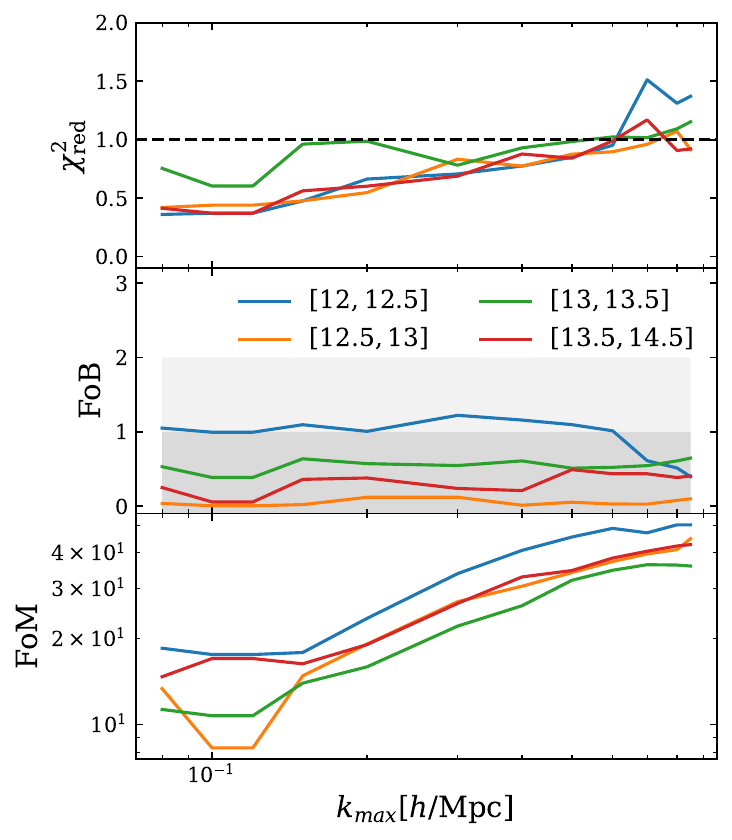}
    \caption{\textit{Upper Panel:} Values of $\chi_{\rm{red}}^2$ obtained from the fits to the multiple halo samples. Different colors indicate the results for the 4 mass bins. A dashed black line marks the value of 1. \textit{Middle Panel:} Figure of bias values as a function of the maximum scale used in the fit. No clear tendency with mass is detected. \textit{Lower Panel:} Figure of merit values as a function of the maximum scale used in the fit. The dependency of this quantity with mass is unclear.}
    \label{fig:chi2_mass_bins}
\end{figure}

Figure \ref{fig:chi2_mass_bins} shows the values of the reduced $\chi^2$ ($\chi^2_{\rm{red}})$, figure of bias (FoB) and figure of merit (FoM) for the fits obtained with HYMALAIA for the power-spectrum multipoles computed from the 4 previously defined halo samples. The fiducial linear alignment bias parameters used to compute the FoB can be consulted in Table \ref{tab:fiducial_parameters}. The first two panels show that the model describes well the halo samples and measures consistent bias parameters until the smallest scales probed, $k= 0.85\,h/$Mpc. In the third panel, one can see that the FoM has an unstable behaviour at the largest scales, but then increases approximately monotonically with the addition of smaller scales beyond $k= 0.2\,h/$Mpc; furthermore, there is no clear behaviour of the FoM with mass. The linear alignment bias of the halo populations increases with the mass, thus increasing the amplitude of the signal and consequently the FoM; however, the number density decreases with mass, thus increasing the level of shape-noise and decreasing the FoM. Since the two effects work against each other, and our mass samples were not chosen to keep one of them fixed, they generate the complex behaviour observed.

\subsubsection{Minimal HYMALAIA}
\label{sec:minimal}

\begin{figure}
	\includegraphics[width=\columnwidth]{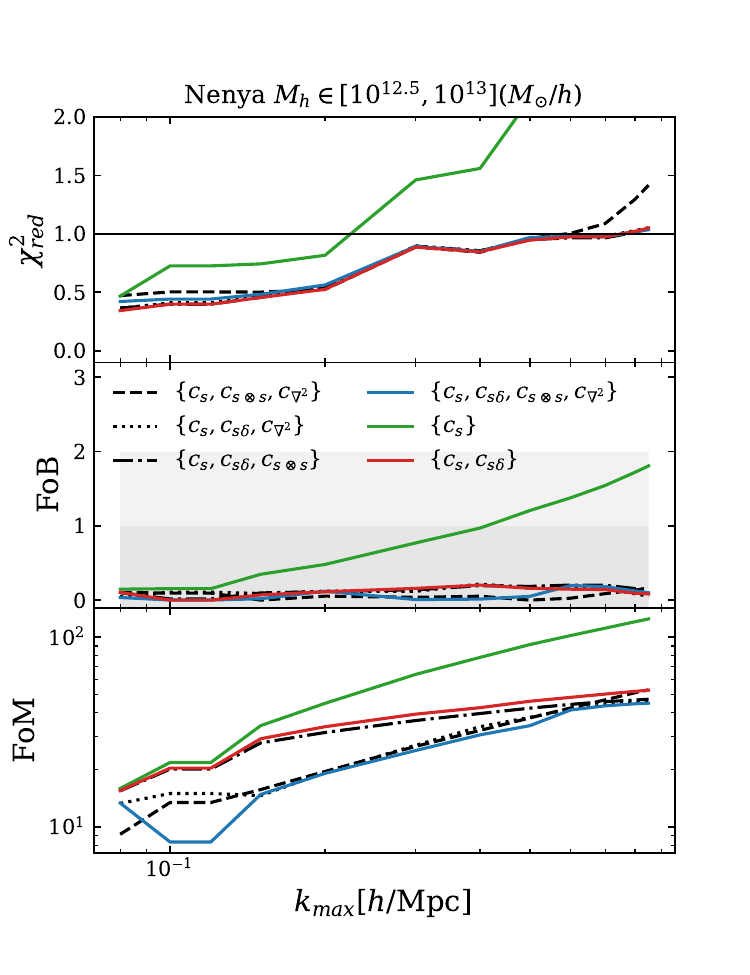}
    \caption{Figure showing the comparison between different cases of HYMALAIA, in which we choose to keep only some of the bias parameters free. Blue lines in this plot represent the full model, with all parameters free. \textit{Top Panel:} Values of $\chi^2$ obtained from the fits to the intermediate mass halo sample. Labels indicate which bias parameters are used in computing each line. \textit{Middle Panel:} Figure of bias values as a function of the maximum scale used in the fit.
    \textit{Lower Panel:} Figure of merit values as a function of the maximum scale used in the fit. }
    \label{fig:chi2_model_terms_compare_kmax}
\end{figure}

Once we have established that HYMALAIA is capable of accurately describing the ensemble of measurements over a wide range of scales, and for several mass-bins, a question naturally arises, whether one can suppress some terms in our model to improve the FoM, while maintaining the accuracy. Figure \ref{fig:chi2_model_terms_compare_kmax} shows the relevant comparison to answer this question, analyzing the relative importance of the terms entering the bias expansion. From the first panel one can see that the terms accompanying $c_s$ and $c_{s\delta}$ are the most important in the expansion, since the model with only these terms fits the data equally well as its more complex counterparts. In the second panel one can see that, except for the model containing only $c_s$, none of the others demonstrate a detectable bias in their recovered values of $c_s$. Finally, the third panel shows that adding more parameters, in particular $c_{\nabla^2}$, can significantly reduce the constraining power of the model. The full picture suggests that the optimal version of the model is the one keeping only $c_s, c_{s\delta}$, since its accuracy is the same as the one of the full model, it gives unbiased constraints on $c_s$, and is the one with the largest FoM among the models that are still accurate at small scales. This picture is preserved qualitatively for the remaining mass-bins and cosmologies. We shall label this reduced version of the model min-HYMALAIA.

\subsubsection{Model Comparison}

\begin{figure}
    \includegraphics[width=\columnwidth]{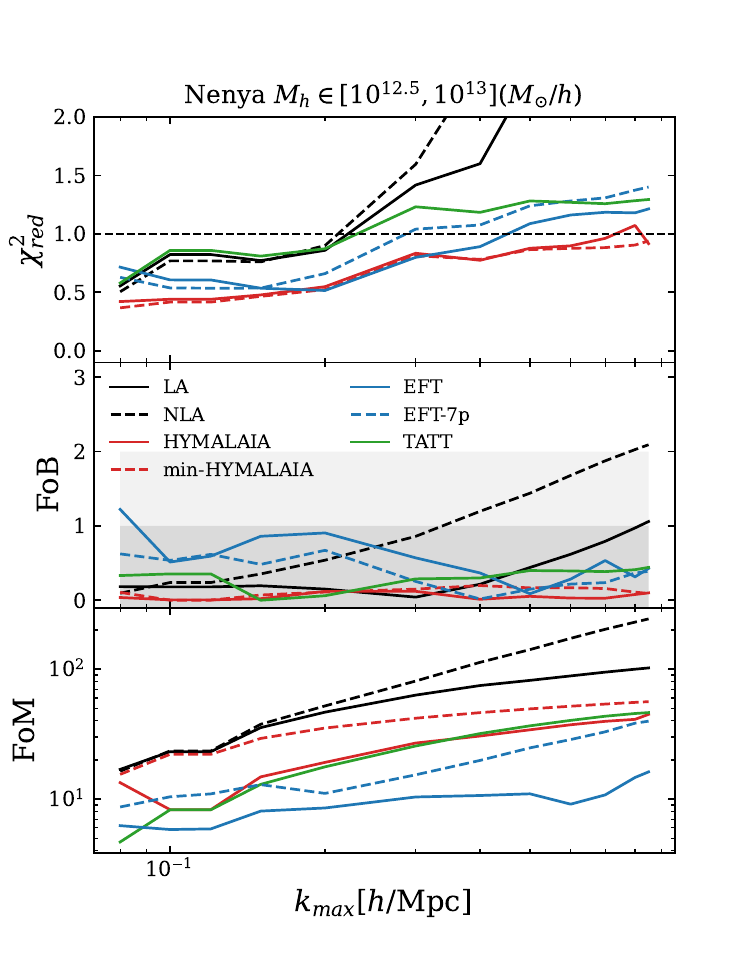}
    \caption{This figure displays comparisons between HYMALAIA and the previously existing models, described in section \ref{sec:Model}. \textit{Top Panel:} Values of reduced $\chi^2$ obtained from the fits to the intermediate-mass halo sample.
    \textit{Middle Panel:} Figure of bias values as a function of the maximum scale used in the fit. \textit{Bottom Panel:} Figure of merit values as a function of the maximum scale used in the fit.
    }
    \label{fig:chi2_la_nla}
\end{figure}

Having established that HYMALAIA and min-HYMALAIA are capable of accurately describing the set of observables over the probed range of scales, we now wish to compare them to other pre-existing models. Figure \ref{fig:chi2_la_nla} shows a comparison of $\chi^2_{\rm{red}}$, FoB and FoM, but only for one mass bin, now varying the model used to fit the data. The first panel shows that the two flavours of HYMALAIA give the best fits to the data among the compared models, with the values for $\chi^2_{\rm{red}}$ well below those of LA, NLA and TATT, and marginally better performances than EFT of IA. It also shows that TATT recovers marginally lower values than the 7-parameter EFT of IA beyond $k\approx0.4\,h/$Mpc. This is likely due to the fact that the implementation of TATT by \cite{Blazek_2019} uses the fully non-linear power spectrum instead of the precise perturbative expressions, so that it extends beyond the 1-loop expansion of the EFTofIA, albeit inconsistently in perturbation theory; \cite{Bakx_2023} finds that the 7-parameter EFTofIA always performs better than the pertubatively-consistent implementation of TATT, thus supporting our finding.

The second panel shows that all the models, with the exception of NLA, recover values of $c_s$ that are less than 1-$\sigma$ distant from the fiducial large-scale value. This is particularly important because one generally expects $c_s$ to be the parameter with the greatest degeneracy with the cosmological parameters of interest, and crucially other systematics such as  photo-$z$s \citep{Fischbacher_2023}, and thus determining it consistently is essential. 

Finally, the third panel displays the values of the figure-of-merit for each model. One can see that HYMALAIA recovers the same information as TATT. Furthermore, at the smallest scales tested, HYMALAIA recovers roughly the same information as LA or NLA on scales $k\approx 0.2\,h/$Mpc, at which these models start to break down. As for min-HYMALAIA, it recovers as much as twice the information extracted by TATT, depending on the scales of interest, while still maintaining lower values of chi-squared over all the probed range of scales.

It is interesting to investigate what are the reasons due to which HYMALAIA recovers lower $\chi^2_{\mathrm{red}}$ and higher FOM than TATT. We argue that these most likely arise due to the following motives:
\begin{enumerate}
    \item HYMALAIA is based on a Lagrangian bias expansion, effectively meaning that higher-order operators than those included are generated by the advection procedure \citep{Schmitz_2018};
    \item the displacements are computed from $N$-body simulations, which give a more accurate description of the non-linear regime than PT.
\end{enumerate}

One final caveat must be taken into account at this point: as mentioned in section \ref{sec:TATT}, the TATT model employed here is based on the public implementation available in \verb|FAST-PT|, which outputs the spectra assuming the Limber approximation; the dependence on $\mu$ can be included exactly for $P_{\delta E}^{(\ell)}$, but not for $P_{EE}^{(\ell)}, P_{BB}^{(\ell)}$. To compute the multipoles we then include an approximate angle dependence to these spectra. Nevertheless, we argue this approximation should be accurate, since \cite{Bakx_2023} compare the full EFTofIA model to a reduced version of it -- equivalent to TATT, but with exact angle dependence -- to a very similar conclusion as the one found in this work.

\subsection{Bias Relations}
\begin{figure}
	\includegraphics[width=\columnwidth]{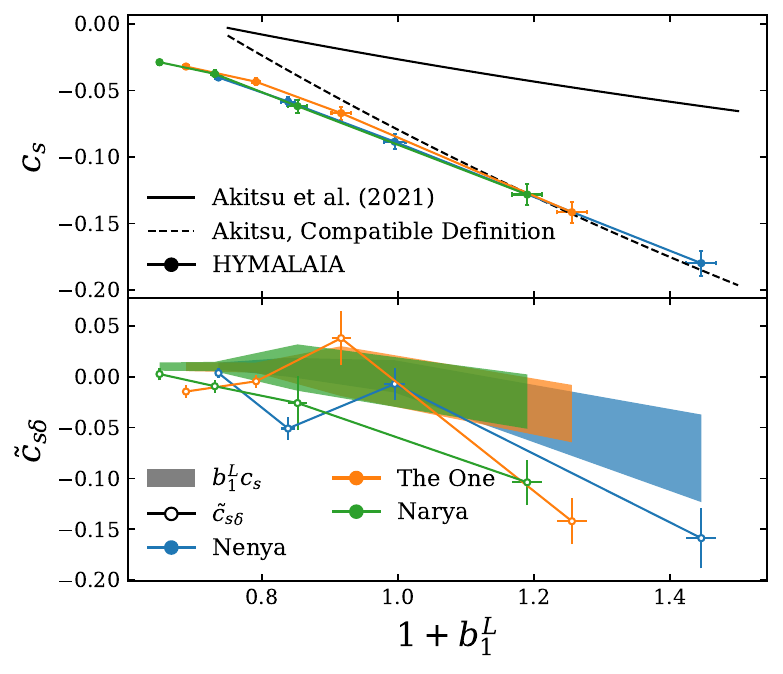}
    \caption{\textit{Top Panel:} Coevolution of the linear IA bias parameter with the first-order density bias parameter of halos. The colorful points show the values obtained directly from applying HYMALAIA to our measurements of the shape power spectra for our three available cosmologies, and the four mass bins within each of them. The black solid line shows the fitting function from \citep{Akitsu_2021}. The black dashed line shows this fitting function after being adapted to be compatible with our definition of the linear alignment bias. The correction is simply to multiply their bias values by $3$; see appendix \ref{sec:bias_compare} for more details, especially equation \eqref{eq:correction_bias}. \textit{Bottom Panel:} Comparison between measured values of $\tilde{c}_{s\delta}$, and values expected from assuming only linear Lagrangian alignment, $\tilde{c}_{s\delta} = c_s b_1$.
    }
    \label{fig:bias_params}
\end{figure}

This sample of halos also allows one to determine a relation between the usual linear halo bias $b_1$ and the linear IA bias, $c_s$. In Figure \ref{fig:bias_params} we show the measurements of these values from the set of simulations previously introduced. This points to a strong relationship between $b_1$ and $c_s$ which furthermore appears to be redshift and cosmology 
independent, in qualitative agreement with the findings of \cite{Akitsu_2021}; a quantitative comparison with the results of that work requires careful analysis of the differences in definition of the bias parameters. Appendix \ref{sec:bias_compare} discusses how one can account for the difference in definition between these works, to make comparable the bias values obtained in the present work, and the ones measured by \cite{Akitsu_2021}. This can be seen in Figure \ref{fig:bias_params}, in which the solid line in the top panel shows the original fitting function from \cite{Akitsu_2021}, and the dashed line is the one made compatible with our definition. We can see this restores concordance between the two measurements, especially at large values of the linear density bias. Notice, however, that the original fitting of \cite{Akitsu_2021} is done over a very large range of values of $b_1^L$, and does not extend to values below $1+b_1^L\approx 0.75$, hence we expect the fit to be much less accurate in this regime, and the comparison at small values of $b_1^L$ must be made with care.

The second pannel shows the comparison of the values of $\tilde{c}_{s\delta}$ measured with HYMALAIA, and the prediction for this measurement in the case where the linear Lagrangian model would be valid. One can see that in general we detect a difference between these two values, giving some indication of a non-zero value of the  $c_{s\delta}$. 

\section{Conclusions}
\label{sec:conclusions}

The results presented in this work show that HYMALAIA is capable of fitting shape power spectra obtained from populations of halos over a wide range of masses, and for three very distinct cosmologies at different redshifts. Particularly remarkable is the comparison of HYMALAIA and TATT, since the two models have the same amount of free parameters in their most complex forms, but the accuracy of the former exceeds that of the latter, even if one restricts HYMALAIA to its optimal case with only three free parameters.

In order to apply the method for the mitigation of IA as a contaminant in WL analyses, it is fundamental to demonstrate that it describes well the signal, so as not to misinterpret part of the apparent alignment signal induced by lensing as being intrinsic. It is also fundamental that it recovers reliable estimates of the bias parameters, particularly $c_s$, since this is expected to be degenerate with cosmological parameters and systematic effects of great importance. The results in the previous sections have robustly shown that the values of the linear alignment bias $c_s$ are consistently recovered, being in excellent agreement with the fiducial values determined using LA at linear scales; besides this, we can see a rough agreement between the values of $c_s$ found in this work, and the ones reported in \citep{Akitsu_2021}, even though the methodologies used are very distinct, and the focus of that work is on much larger values of the linear Lagrangian bias, so that the comparison at low $b_1^L$ is very uncertain.

Finally, it is also worthy of notice that, even though other models such as the EFTofIA can extend over a similar range of scales while remaining accurate and unbiased, the small number of parameters of HYMALAIA causes it to recover considerably more information from the same dataset. This can be quantitatively evaluated by the FoM, and results presented previously showed that HYMALAIA has the highest FoM among the methods capable of extending beyond $k\approx 0.2\,h/$Mpc. Restricting to min-HYMALAIA makes this picture even more optimistic, with FoM values that are $\sim 20\%$ larger than those for HYMALAIA at the smallest scales probed, even if tests must still be made to understand whether min-HYMALAIA provides a good description of galaxy shape correlators, as it has been shown to do for halos.

Since HYMALAIA is based on $N$-Body simulation results, its application to e.g. mitigation of IA in cosmic-shear surveys, would depend on building an emulator for its basis spectra as a function of cosmology, which is beyond the scope of this work. However, even without building such emulators, one can consider using the model for infusing the IA signal into simulations and exploring its effect on higher-order statistics \citep{Harnois-Deraps_2022, vanalfen2023empirical}. 

\section*{Acknowledgements}


The authors thank Sergio Contreras for useful discussions and technical support. We also thank Zvonimir Vlah for useful discussions at the start of this project, and comments on earlier versions of this manuscript.

The authors acknowledge the support of the E.R.C. grant 716151 (BACCO). REA acknowledges the support of the Project of excellence Prometeo/2020/085
from the Conselleria d’Innovació, Universitats, Ciéncia i Societat
Digital de la Generalitat Valenciana, and of the project PID2021-
128338NB-I00 from the Spanish Ministry of Science.

This publication is part of the project ``A rising tide: Galaxy intrinsic alignments as a new probe of cosmology and galaxy evolution'' (with project number VI.Vidi.203.011) of the Talent
programme Vidi which is (partly) financed by the Dutch Research Council (NWO). For the purpose of open access, a CC BY public copyright
license is applied to any Author Accepted Manuscript
version arising from this submission.

This work is also part of the Delta ITP consortium, a program of the Netherlands Organisation for
Scientific Research (NWO) that is funded by the Dutch Ministry of Education, Culture and Science (OCW). We also thank the Lorentz Center for helping us start this project during the workshop ``hol-IA: a holistic approach to galaxy intrinsic alignments'' that took place in March 2023\footnote{\tiny{\url{https://www.lorentzcenter.nl/hol-ia-a-holistic-approach-to-galaxy-intrinsic-alignments.html}}}.

\section*{Data Availability}

The data underlying this article will be shared on reasonable
request to the corresponding author.



\bibliographystyle{mnras}
\bibliography{hybrid_IA} 



\appendix

\section{Comparing Bias Parameters}
\label{sec:bias_compare}

As previously stated in the text, the values of the bias parameters obtained in this work cannot be directly compared to those reported in \cite{Akitsu_2021, Akitsu_2023b}; let us explore in more detail why that is. In these works, their parameter of interest, $b_K$, is defined via the expression
\begin{equation}
    \frac{1}{S_0}\left[S_{ij} - \frac{1}{3}S_0\delta^K_{ij}\right] = b_K s_{ij}, 
\end{equation}
in which $S_{ij}$ is the 3D shape tensor, $S_0$ is the trace of this tensor, and $s_{ij}$ is the 3D traceless tidal tensor. From this expression these authors then define the following quantities \citep{St_cker_2021}
\begin{equation}
    \begin{split}
        S_p & \equiv S_{33} - \frac{S_{11} + S_{22}}{2}\\
        S_e & \equiv \frac{S_{11} - S_{22}}{2},
    \end{split}
\end{equation}
and from these then extract the bias parameter
\begin{equation}
    \begin{split}
        b_K & = -\frac{S_p(\Delta_p=+\epsilon) - S_p(\Delta_p=-\epsilon)}{2\epsilon D(z) S_0}\\
        b_K & = -\frac{S_e(\Delta_e=+\epsilon) - S_e(\Delta_e=-\epsilon)}{2\epsilon D(z) S_0}.
    \end{split}
\end{equation}
In this work, however, we use a different method for measuring the linear bias. We define the 2D projected shape tensor as being the restriction of $S_{ij}$ to a two-dimensional sub-space; if we choose the projection direction as being $\hat{z}$, this reads
\begin{equation}
    S^{2D}_{ij} = S_{i\leq2, j\leq2}.
\end{equation}
From this shape tensor one can then build the ellipticity fields as in equation (\ref{eq:ellipticity_definition}), which can be written in this new notation as
\begin{equation}
    \begin{split}
        \epsilon_1 & = \frac{S^{2D}_{xx} - S^{2D}_{yy}}{2S_0^{2D}},\\
        \epsilon_2 & = \frac{S^{2D}_{xy}}{S_0^{2D}}, 
    \end{split}
\end{equation}
corresponding to 
\begin{equation}
    \frac{1}{S_0^{2D}}\left[S_{ij}^{2D} - \frac{1}{2}S_0^{2D}\delta^K_{ij}\right] = \frac{c_s}{2}s_{ij}.
\end{equation}
Notice that the normalization of these ellipticities is given by $S_0^{2D}$ instead of the full trace; this can be written as
\begin{equation}
    \langle S_0^{2D} \rangle = \frac{2S_0}{3},
\end{equation}
and there is a factor of $2$ difference in the definitions of the bias parameters. Therefore, our bias parameter will be related to the one of \cite{Akitsu_2021, Akitsu_2023b} by
\begin{equation}
    c_s = 3 b_K.
    \label{eq:correction_bias}
\end{equation}

\section{Notation}

In this work we will employ three closely connected quantities, which can therefore be confused. To avoid this from happening, we will clarify their definitions and make explicit their relations to each other. The shape tensor, $S_{ij}$ has already been defined in equation \eqref{eq:shape_tensor}. It is closely related to $g^{\mathrm{2D}}_{ij}$, which we label the reduced shape tensor, and which is defined by
\begin{equation}
    g^{\mathrm{2D}}_{ij}(\mathbf{x}) = \frac{1}{S_0^{\mathrm{2D}}}\left( S_{ij} - \frac{1}{3}\delta^K_{ij}S^{2\rm{D}}_0 \right),
\end{equation}
in which we have already restricted this tensor to the 2D subspace, corresponding to the fact that galaxy shapes are a projected quantity. Finally, $\gamma_{ij}$ is the symbol we will use for shear, meaning that the intrinsic shear is connected to the reduced shape tensor by
\begin{equation}
    \gamma^I(\mathbf{x}) = \frac{1}{2}\left( g^{\mathrm{2D}}_{11}-g^{\mathrm{2D}}_{22}, 2g^{\mathrm{2D}}_{12} \right).
\end{equation}

\section{Fiducial Parameter Estimates}

The fiducial values of $c_s$ and $A_{SN}$, measured according to what was described in \ref{sec:covariance_matrix} are summarised in table \ref{tab:fiducial_parameters}, corresponding to the halo populations extracted from simulations with Nenya, Narya, and TheOne cosmologies respectively.

\setlength{\tabcolsep}{3pt} 

\begin{table}
    \centering
        \begin{tabular}{|c|c|c|c|c|}
        \hline
        & & Nenya & Narya & TheOne\\
        \hline
        \multirow{4}{*}{$c_s$} & $M_1$ & $-0.0352\pm 0.0047$ & $-0.0286\pm 0.0067$ & $-0.0319\pm 0.0049$ \\
        & $M_2 $& $-0.057\pm 0.013$ & $-0.041\pm 0.016$ & $-0.0399\pm 0.0093$\\
        & $M_3$ & $-0.113\pm 0.042$ & $-0.063\pm 0.030$ & $-0.067\pm 0.019$\\
        & $M_4$ & $-0.202\pm 0.057$ & $-0.097\pm 0.023$ & $-0.125\pm 0.052$\\
        \hline
        \multirow{4}{*}{$A_{SN}$} & $M_1$ & $5.459\pm 0.0093$ & $4.142\pm 0.090$ & $5.170\pm 0.072$ \\
        & $M_2 $& $16.71\pm 0.37$ & $11.69\pm 0.16$ & $15.28\pm 0.11$\\
        & $M_3$ & $56.1\pm 1.1$ & $36.91\pm 0.5$ & $48.8\pm 0.4$\\
        & $M_4$ & $165.9\pm 2.8$ & $102.6\pm 1.6$ & $138.6\pm 1.8$\\
        \hline
        \end{tabular}
    \caption{Fiducial model parameters for each of the halo populations extracted from the mid-sized Nenya simulation. The mass ranges of these samples are defined in Table  \ref{tab:mass_samples}.}
    \label{tab:fiducial_parameters}
\end{table}

\section{Priors}

In Bayesian analyses, one needs to define prior probability density functions for the parameters, to encode the knowledge one may have on the behaviour of that quantity, prior to any analysis. In this case, since we have no expectation on the behaviour of these bias parameters, we choose to set flat priors over a wide range of values. To know whether these priors are wide enough, we observe the recovered posterior distribution, and check whether it is encountering the prior in a region at which it has high-probability values. If this is the case, we then enlarge the prior and check again until this no longer takes place. The final ranges used in the analyses presented in this work can be seen in Table \ref{tab:priors}.

\begin{table}
 \caption{Ranges of the flat priors employed in the Bayesian analysis of simulation data in the light of the EFTofIA model. These priors are designed to be sufficiently broad to not interfere with the behaviour of the chains in exploring the posterior distribution; this is done in an empirical fashion, by running the MultiNest algorithm with one set of priors, and updating them as necessary, in case a region with relevant probability density has been excluded.}
 \begin{tabular}{cc}
  \hline
   & Prior Range \\
   \hline
   $c_s$ & [-1,0] \\
   $c_{2,1}$ & [-3,3]  \\
   $c_{2,2}$ & [-3,3]  \\
   $c_{2,3}$ & [-3,3]  \\
   $c_{3,1}$ & [-3,3]  \\
   $c_{3,2}$ & [-3,3]  \\
   $c_{\nabla^2}$ & [-3,3] \\
   $b_{\nabla^2}$ & [-5,5]    \\
   $A_{SN}$ & [0,3000] \\
  \hline
 \end{tabular}
 \label{tab:priors}
\end{table}

\section{Covariance Matrix Modeling}
\label{sec:cov_matrix}

\begin{figure}
    \centering
    \includegraphics[width=\columnwidth]{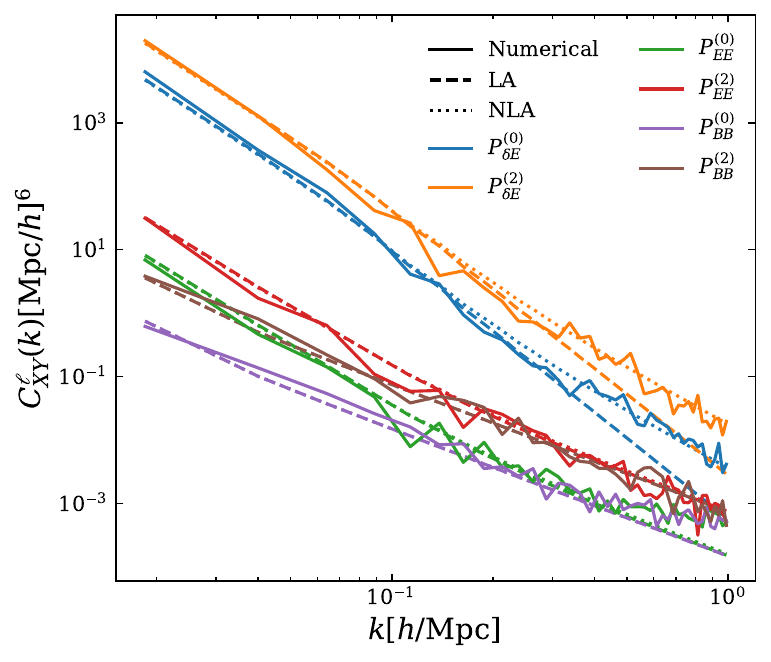}
    \caption{Comparison of the diagonal terms of the covariance matrix computed analytically using LA or NLA, and the numerical one, computed by \citep{Kurita_2020} from the set of simulations of the Dark Quest project \citep{Nishimichi_2019}. This result is for halos of mass in the range $\log_{10}(M/M_{\odot}) \in [12.5,13]$; Notice that in this figure we are displaying only the auto covariance between spectra, while equation \eqref{eq:cov_model} also makes predictions for the cross-covariances between different type of spectra and multipoles. }
    \label{fig:cov_mat_compare}
\end{figure}

To validate the assumption made in equation \eqref{eq:cov_model}, and the use of the NLA model in computing the covariance matrix, we compare the results from this calculation to a numerical one, which should accurate, even though noisier. The numerical estimate is obtained from the measured power spectra in \cite{Kurita_2020} from a set of 20 simulations which are part of Dark Quest simulation suite \citep{Nishimichi_2019}. From these one can measure the shape power spectra, and then compute the numerical covariance
\begin{equation}
    \begin{split}
        C_{\alpha,\beta}^{\ell,\ell'}(k_i,k_j) = \frac{1}{N-1}\sum_{n=1}^{N} \Big( P^{(\ell)}_{\alpha, n}(k_i) - &  \big\langle{P}^{(\ell)}_\alpha(k_i)\big\rangle\Big) \\
        &\Big(P^{(\ell')}_{\beta, n}(k_j) - \big\langle P^{(\ell')}_\beta(k_j) \big\rangle \Big),
    \end{split}
\end{equation}
with $N=20$. In Figure \ref{fig:cov_mat_compare} we can see a comparison of the analytical covariance calculation assuming different models for the shape power spectra, and the numerical calculations. From this we conclude that assuming NLA in this case is a better approximation than using LA.

\begin{figure}
    \centering
    \includegraphics[width=\columnwidth]{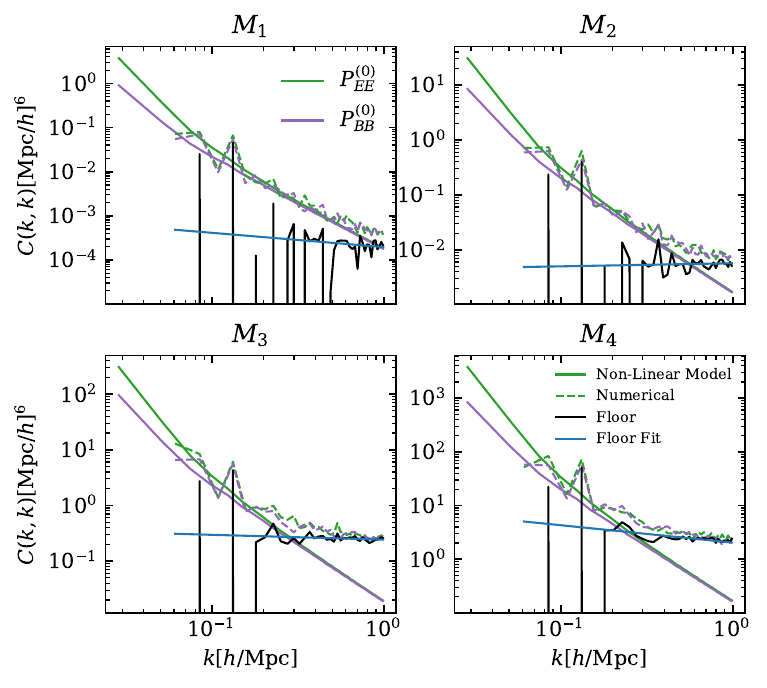}
    \caption{Comparison between the diagonal of the covariances computed from the analytic approximation of equation \eqref{eq:cov_model} and using the jacknife method. One can see a generally good agreement for intermediate scales $k\approx0.1\,h/$Mpc, and large differences appearing as one goes to smaller scales. Black solid lines show the difference between the two, and solid blue lines show the model fitted to these values.}
    \label{fig:jacknife_cov}
\end{figure}

However, as stated in the text, the approximation of equation \eqref{eq:cov_model} does not reproduce well the behaviour of the diagonal of the auto-covariance for the spectra $P_{EE}^{(0)}$ and $P_{BB}^{(0)}$. Therefore, we have used the jackknife technique to robustly estimate the covariance at small scales. Each of the simulations in Table \ref{tab:cosmologies} with $V=(512\,h^{-1}$Mpc$)^3$ were divided into 64 smaller simulations of volume $V=(128\,h^{-1}$Mpc$)^3$; each of these simulations will give us a measurement of the shape-power spectra, and we can then compute the auto-covariance from the ensemble of spectra
\begin{equation}
    \begin{split}
        C^{(\ell)}_{XY}(k,k') = \frac{1}{64-1}\sum_{n=1}^{64} & \left( P^{(\ell)}_{XY, n}(k) - \Big\langle P^{(\ell)}_{XY}(k) \Big \rangle  \right) \\
        & \left( P^{(\ell)}_{XY, n}(k') - \Big\langle P^{(\ell)}_{XY}(k') \Big\rangle  \right).
    \end{split}
\end{equation}
We will denote the difference between the numerical estimate and the analytical approximation as the \textit{floor} of the covariance, which can be seen represented by the black lines in Figure \ref{fig:jacknife_cov}. We parametrize the floor for the different mass-bins as a power-law
\begin{equation}
    F(k) = Ak^B,
\end{equation}
and find the best-fitting values of $A$ and $B$ for each mass-bin. This model computed with the best-fitting parameters is represented by the blue lines in Figure \ref{fig:jacknife_cov}.

\begin{figure*}
    \centering
    \includegraphics[width=0.7\textwidth]{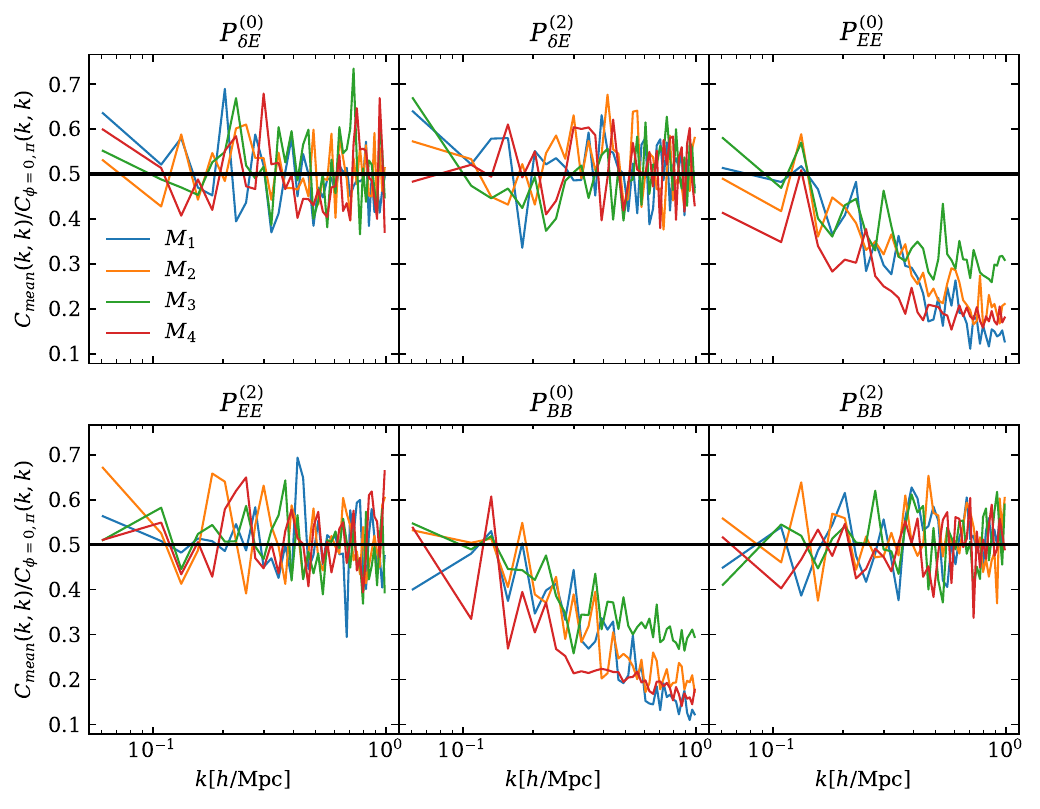}
    \caption{Ratio of the diagonal of the jacknife covariances computed either from the mean of the two paired simulations, or from one of them alone. Values of $0.5$ indicate the scenario where the deviations from the ensemble average in the two simulations are completely uncorrelated, thus functioning essentially as two random realizations. The cases when this ratio falls below $0.5$ indicate the case where the deviations are anti-correlated, and thus cancel one another when averaging over the two phases.}
    \label{fig:pairing_jack_cov}
\end{figure*}

The estimation of the covariance matrices with the jacknife method also allows us to probe the effect of F\&P on the shape-power spectra. These results can be seen in Figure \ref{fig:pairing_jack_cov} and are discussed in section \ref{sec:simulations}.

\bsp	
\label{lastpage}
\end{document}